\documentclass[english,manuscript,A4,english,aps]{revtex4-1}
\usepackage[T1]{fontenc}
\usepackage[latin9]{inputenc}
\usepackage{color}
\usepackage{babel}
\usepackage{textcomp}
\usepackage{amstext}
\usepackage{amssymb}
\usepackage{graphicx}
\usepackage{esint}
\usepackage[unicode=true,pdfusetitle,
 bookmarks=true,bookmarksnumbered=false,bookmarksopen=false,
 breaklinks=false,pdfborder={0 0 0},pdfborderstyle={},backref=false,colorlinks=true]
 {hyperref}
\hypersetup{
 linkcolor=blue, citecolor=cyan}

\makeatletter

\providecommand{\tabularnewline}{\\}

\usepackage{xcolor}

\makeatother

\begin{document}
\title{Anomalous magnetohydrodynamics with temperature-dependent electric
conductivity and application to the global polarization }
\author{Hao-Hao Peng}
\affiliation{Department of Modern Physics, University of Science and Technology
of China, Hefei 230026, China}
\author{Sihao Wu}
\affiliation{Department of Modern Physics, University of Science and Technology
of China, Hefei 230026, China}
\author{Ren-jie Wang}
\email{wrjn@mail.ustc.edu.cn}

\affiliation{Department of Modern Physics, University of Science and Technology
of China, Hefei 230026, China}
\author{Duan She}
\email{sheduan@ustc.edu.cn}

\affiliation{Department of Modern Physics, University of Science and Technology
of China, Hefei 230026, China}
\author{Shi Pu}
\email{shipu@ustc.edu.cn}

\affiliation{Department of Modern Physics, University of Science and Technology
of China, Hefei 230026, China}
\begin{abstract}
We have derived the solutions of the relativistic anomalous magnetohydrodynamics
with longitudinal Bjorken boost invariance and transverse electromagnetic
fields in the presence of temperature or energy density dependent
electric conductivity. We consider the equations of states in a high
temperature limit or in a high chiral chemical potential limit. We
obtain both perturbative analytic solutions up to the order of $\hbar$
and numerical solutions in our configurations of initial electromagnetic
fields and Bjorken flow velocity. Our results show that the temperature
or energy density dependent electric conductivity plays an important
role to the decaying of the energy density and electromagnetic fields.
We also implement our results to the splitting of global polarization
for $\Lambda$ and $\overline{\Lambda}$ hyperons induced by the magnetic
fields. Our results for the splitting of global polarization disagree
with the experimental data in low energy collisions, which implies
that the contribution from gradient of chemical potential may dominate
in the low energy collisions.
\end{abstract}
\maketitle

\section{Introduction}


Recently, strong electromagnetic (EM) fields (of order $10^{17}-10^{18}$Gauss
\citep{Skokov:2009qp,Bzdak:2011yy,Deng:2012pc,Roy:2015coa}) generated
in the relativistic heavy ion collisions provide a platform to study
the novel quantum transport phenomena and non-linear quantum electrodynamics. 


The strong magnetic fields can induce the chiral magnetic effect (CME)
\citep{Vilenkin:1980fu,Kharzeev:2007tn,kharzeev2008effects,Fukushima:2008xe}
associated with the chiral anomaly, chiral separation effect \citep{Aghababaie:2003iw}
and chiral magnetic wave \citep{Kharzeev:2010gd,Burnier:2011bf}.
While the electric fields can cause chiral electric separation effect
\citep{Huang:2013iia,Pu:2014cwa,Jiang:2014ura,Pu:2014fva} and other
novel nonlinear EM responses in chiral systems \citep{Pu:2014fva,Chen:2016xtg,Gorbar:2016qfh,Gorbar:2016sey,Gorbar:2016ygi,Gorbar:2017cwv}.
These chiral transport phenomena are naturally connected to the Schwinger
mechanism \citep{Fukushima:2010vw,Warringa:2012bq,Copinger:2018ftr,Copinger:2020nyx,Copinger:2022jgg}.
More discussions and references can be found in recent reviews \citep{Kharzeev:2012ph,Kharzeev:2015znc,Liao:2014ava,Miransky:2015ava,Huang:2015oca,Fukushima:2018grm,Bzdak:2019pkr,Zhao:2019hta,Gao:2020vbh,Shovkovy:2021yyw}.
To understand and describe these chiral transport, the quantum kinetic
theory or named Wigner function approaches, as an microscopic effective
theory has been widely discussed \citep{Gao:2012ix,Son:2012zy,Chen:2012ca,Stephanov:2012ki,Manuel:2013zaa,Chen:2013dca,Manuel:2014dza,Chen:2014cla,Chen:2015gta,Hidaka:2016yjf,Mueller:2017lzw,Hidaka:2017auj,Gao:2017gfq,Liu:2018xip,Hidaka:2018ekt,Huang:2018wdl,Hidaka:2018mel,Gao:2018wmr,Gao:2019znl,Wang:2019moi,Li:2019qkf,Weickgenannt:2019dks,Hattori:2019ahi,Lin:2019ytz,Lin:2019fqo,Gao:2019zhk,Weickgenannt:2020aaf,Weickgenannt:2020sit,Liu:2020flb,Huang:2020wrr,Yamamoto:2020zrs,Yang:2020hri,Wang:2021qnt,Sheng:2021kfc,Luo:2021uog,Weickgenannt:2021cuo,Hidaka:2022dmn,Fang:2022ttm}
(also see Ref. \citep{Hidaka:2022dmn,Gao:2020vbh,Gao:2020pfu} for
the recent reviews on quantum kinetic theory). 


It is still challenging to search the CME signal in the relativistic
heavy ion collision at Relativistic Heavy Ion Collider (RHIC) \citep{STAR:2009wot,STAR:2009tro,Wang:2012qs,STAR:2013ksd,STAR:2013zgu,STAR:2014uiw,Tribedy:2017hwn,STAR:2019xzd}
and Large Hadron Collider (LHC) \citep{ALICE:2012nhw,CMS:2016wfo,CMS:2017lrw}.
Although many possible correlators or observables have been proposed
\citep{STAR:2013ksd}, the background contributions from, e.g., the
transverse momentum conservation \citep{Pratt:2010zn,Bzdak:2012ia,Ajitanand:2010rc,Magdy:2017yje,Tang:2019pbl},
local charge conservation \citep{Schlichting:2010qia} and non-flow
correlations \citep{Xu:2017zcn,Xu:2020sln}, cannot be neglected.
Nevertheless, the recent isobar collisions \citep{Voloshin:2010ut}
in STAR measurements \citep{STAR:2021mii} do not observe the CME
signature that satisfies the predefined criteria. It requires the
future systematical studies on CME.


Another interesting phenomena related to the chiral transport and
EM fields is the global and local polarization of $\Lambda$ and $\overline{\Lambda}$
hyperons \citep{Liang:2004ph,Liang:2004xn,Gao:2007bc}. The global
polarization has been measured by the STAR experiments \citep{STAR:2017ckg}
and been widely studied in several models \citep{Karpenko:2016jyx,Li:2017slc,Xie:2017upb,Sun:2017xhx,Shi:2017wpk,Wei:2018zfb,Wu:2019eyi,Wu:2020yiz,Fu:2020oxj,Ryu:2021lnx,Wu:2022mkr}
(also see recent reviews \citep{Wang:2017jpl,Becattini:2020ngo,Becattini:2020sww,Gao:2020vbh}).
Meanwhile, the local spin polarization, in Au+Au collisions at $\sqrt{s_{NN}}=$
200 GeV has been measured \citep{STAR:2019erd} and studied
\citep{Liu:2020dxg,Liu:2021uhn,Becattini:2021suc,Yi:2021ryh,Ryu:2021lnx,Yi:2021unq,Wu:2022mkr}.
Several studies have pointed out that the splitting of global polarization
of $\Lambda$ and $\overline{\Lambda}$ hyperons may be induced by
the EM fields \citep{Muller:2018ibh,Guo:2019joy,Buzzegoli:2022qrr,Xu:2022hql}
and gradient of baryon chemical potential \citep{Ryu:2021lnx,Fu:2022myl,Wu:2022mkr}.
Therefore, the further studies on the EM fields and their evolution
are also required in this field.


On the other hand, strong EM fields make the studying of the non-linear
electrodynamics be possible. The light-by-light scattering \citep{ATLAS:2017fur},
matter generation directly from photons \citep{STAR:2019wlg,Zha:2018tlq},
the vacuum birefringence \citep{Hattori:2012je,Hattori:2012ny,Hattori:2020htm,STAR:2019wlg,Hattori:2022uzp,Adler:1971wn}
have been measured. People find that the EM fields can be considered
as the real photons approximations. Therefore, the lepton pair photoproduction
in peripheral and ultra-peripheral collisions have been comprehensively
studied at both experimental \citep{ATLAS:2018pfw,STAR:2018ldd,STAR:2019wlg,ALICE:2022hvk}
and theoretical sides, e.g., by the generalized equivalent photon approximation
or the calculations based on QED models in the background field approaches
\citep{Vidovic:1992ik,Hencken:1994my,Hencken:2004td,Zha:2018tlq,Zha:2018ywo,Brandenburg:2020ozx,Brandenburg:2021lnj,Li:2019sin},
the framework based on the factorization theorem \citep{Klein:2018fmp,Klein:2020jom,Li:2019yzy,Xiao:2020ddm}
and the calculation based on QED with the assumption of wave-packet
\citep{Wang:2021kxm,Wang:2022gkd,Lin:2022flv}.


As known, the relativistic hydrodynamics \citep{Kolb:2000sd,Kolb:2003dz,Hama:2004rr,Huovinen:2006jp,Ollitrault:2007du,Teaney:2003kp,Lacey:2006bc,Gale:2013da}
is an macroscopic effective theory for the relativistic many body
systems. To learn the chiral transport phenomena, the relativistic
anomalous magentohydrodynamics (MHD), which are the ordinary relativistic
hydrodynamics coupled to the Maxwell's equation in the presence of
chiral anomaly and CME, has been widely discussed \citep{Pu:2016ayh,Roy:2015kma,Pu:2016bxy,Pu:2016rdq,Siddique:2019gqh,Wang:2020qpx,shokri2018evolution,Shokri:2017xxn,Inghirami:2016iru,Biswas:2020rps}.
Meanwhile, the simulations from the anomalous-viscous fluid dynamics
with a given background EM fields are presented in Refs. \citep{Jiang:2016wve,Shi:2017cpu,Shi:2017ucn}.
Similarly, to describe the polarization effects, the spin hydrodynamics
have been built \citep{Montenegro:2017lvf,Montenegro:2017rbu,Florkowski:2017dyn,Florkowski:2017ruc,Florkowski:2018myy,Florkowski:2018fap,Becattini:2018duy,Yang:2018lew,Florkowski:2018ahw,Florkowski:2019qdp,Hattori:2019lfp,Florkowski:2019voj,Bhadury:2020puc,Shi:2020qrx,Fukushima:2020qta,Fukushima:2020ucl,Li:2020eon,Singh:2020rht,She:2021lhe,Gallegos:2021bzp,Hongo:2021ona,Florkowski:2021wvk,Wang:2021ngp,Wang:2021wqq,Bhadury:2022ulr,Biswas:2022bht,Cao:2022aku}
(also see the recent reviews \citep{Wang:2017jpl,Florkowski:2018fap,Becattini:2020ngo,Becattini:2020sww,Gao:2020vbh,Liu:2020ymh}).


In previous studies \citep{Siddique:2019gqh,Wang:2020qpx} by some
of us, we have investigated the relativistic anomalous MHD with a
longitudinal boost invariant Bjorken flow and transverse EM fields.
For simplicity, we assume the electric conductivity is a constant.
Nevertheless, the recent numerical simulations from relativistic Boltzmann
equations show that the electric conductivity changes with proper
time \citep{Yan:2021zjc,Zhang:2019uor,Zhang:2022lje,Wang:2021oqq}.
To consider these effects, we need to consider a temperature dependent
electric conductivity, also see the early studies from lattice QCD
\citep{Aarts:2007wj,Ding:2010ga,Tuchin:2013ie} and holographic models
\citep{Pu:2014cwa,Pu:2014fva}. We will solve the anomalous MHD in
a longitudinal boost invariant Bjorken flow with a temperature or
energy density dependent electric conductivity. Meanwhile, we will
implement our results to the splitting of global polarization for
the $\Lambda$ and $\overline{\Lambda}$ hyperons. 

The structure of this work is as follows. In Sec. \ref{sec:2 Anomalous-magnetohydrodynamics},
we briefly review the relativistic anomalous MHD. Then, we consider
two kinds of equations of state, which are the system in a high temperature
or high chiral chemical potential limits and take the system in a
Bjorken flow in Sec. \ref{sec:equations}. We present the perturbative
analytic solutions and numerical solutions with our configurations
of initial EM fields in high temperature and high chiral chemical
potential limits in Sec. \ref{sec:solution_HT} and \ref{sec:solution_HC},
respectively. We implement our results to the splitting of global
polarization of $\Lambda$ and $\overline{\Lambda}$ hyperons in Sec.
\ref{sec:magnetic-field-induced} and summarize the work in Sec. \ref{sec:4 Discussion-and-Summary}.

Throughout this work, we choose the metric $g_{\mu\nu}=\mathrm{diag}\{+,-,-,-\}$
and Levi-Civita tensor $\epsilon^{\mu\nu\rho\sigma}$ satisfying $\epsilon^{0123}=-\epsilon_{0123}=+1$.
Note that $\epsilon^{\mu\nu\alpha\beta}\epsilon_{\mu\nu\rho\sigma}=-2!(g_{\rho}^{\alpha}g_{\sigma}^{\beta}-g_{\sigma}^{\alpha}g_{\rho}^{\beta})$.
The fluid velocity $u^{\mu}=\gamma(1,\mathbf{v})$ with $\gamma$
being Lorentz factor satisfies $u^{\mu}u_{\mu}=1$ and $\Delta^{\mu\nu}=g^{\mu\nu}-u^{\mu}u^{\nu}$
is the orthogonal projector to the fluid four-velocity $u^{\mu}$. 

\section{Anomalous magnetohydrodynamics \label{sec:2 Anomalous-magnetohydrodynamics}}

In this section, we give a brief review for the relativistic anomalous
MHD. The MHD equations are the conservation equations coupled to the
Maxwell's equations (see, e.g., Refs. \citep{gedalin1995generally,caldarelli2009dyonic,Huang:2009ue,Pu:2016ayh,Roy:2015kma,Pu:2016bxy,Pu:2016rdq,Siddique:2019gqh,Wang:2020qpx}).
The energy-momentum conservation equation is, 
\begin{equation}
\partial_{\mu}T^{\mu\nu}=0,\label{eq:EMT-01}
\end{equation}
where the energy-momentum tensor $T^{\mu\nu}$ can be decomposed as
two parts,
\begin{equation}
T^{\mu\nu}=T_{F}^{\mu\nu}+T_{EM}^{\mu\nu}.\label{eq:total-emt}
\end{equation}
Here, $T_{F}^{\mu\nu}$ stands for the energy-momentum tensor of the
medium and is usually written as in the Landau frame, 
\begin{equation}
T_{F}^{\mu\nu}=\varepsilon u^{\mu}u^{\nu}\text{\textminus}(p+\Pi)\Delta^{\mu\nu}+\pi^{\mu\nu},\label{eq:Fuild_Tensor}
\end{equation}
where $\varepsilon,p,\Pi,\pi^{\mu\nu}$ denote the energy density,
pressure, bulk viscosity pressure and shear viscous tensor, respectively.
The $T_{EM}^{\mu\nu}$ is the energy and momentum given by the EM
fields, i.e. 
\begin{equation}
T_{EM}^{\mu\nu}=-F^{\mu\lambda}F_{\;\lambda}^{\nu}+\frac{1}{4}g^{\mu\nu}F^{\rho\sigma}F_{\rho\sigma}.\label{eq:EM_Tensor}
\end{equation}
In the relativistic hydrodynamics, we usually rewrite the EM tensor
$F^{\mu\nu}$ as, 
\begin{equation}
F^{\mu\nu}=E^{\mu}u^{\nu}-E^{\nu}u^{\mu}+\epsilon^{\mu\nu\alpha\beta}u_{\alpha}B_{\beta},\label{eq:F_01}
\end{equation}
where we introduce the four vector form of electric and magnetic fields,
\begin{equation}
E^{\mu}=F^{\mu\nu}u_{\nu},\;B^{\mu}=\frac{1}{2}\epsilon^{\mu\nu\alpha\beta}u_{\nu}F_{\alpha\beta}.\label{eq:EB_def}
\end{equation}
Note that, by definition (\ref{eq:EB_def}), we find that $u^{\mu}E_{\mu}=0$
and $u^{\mu}B_{\mu}=0$. For convenience, we also define 
\begin{eqnarray}
E & = & \sqrt{-E^{\mu}E_{\mu}},\nonumber \\
B & = & \sqrt{-B^{\mu}B_{\mu}}.
\end{eqnarray}

Besides the energy and momentum conservation, we have the charge (or
vector current) $j_{e}^{\mu}$ conservation equations and the anomalous
equations for the axial current (or chiral current) $j_{5}^{\mu}$,
\begin{eqnarray}
\partial_{\mu}j_{e}^{\mu} & = & 0,\nonumber \\
\partial_{\mu}j_{5}^{\mu} & = & -e^{2}CE^{\mu}B_{\mu}.\label{eq:conservation}
\end{eqnarray}
Here, $C=\hbar/(2\pi^{2})$ is the chiral anomaly coefficient \citep{Son:2009tf,Son:2012bg,Pu:2010as,Pu:2012wn,Siddique:2019gqh,Wang:2020qpx}.
The constitution equations for the currents are, 
\begin{eqnarray}
j_{e}^{\mu} & = & n_{e}u^{\mu}+\sigma E^{\mu}+\xi B^{\mu}+\xi_{\omega}\omega^{\mu}+\nu^{\mu},\nonumber \\
j_{5}^{\mu} & = & n_{5}u^{\mu}+\sigma_{5}E^{\mu}+\xi_{5}B^{\mu}+\xi_{5\omega}\omega^{\mu}+\nu_{5}^{\mu}.\label{eq:current_02}
\end{eqnarray}
Here, $n_{e}$ and $n_{5}$ are the electric and chiral charge density,
respectively. The $\nu^{\mu},\nu_{5}^{\mu}$ are the heat conducting
flow for the charge and chiral currents, respectively. The transport
coefficients for conducting flows induced by the electric fields $\sigma,\sigma_{5}$
are computed by Ref. \citep{Huang:2013iia,Pu:2014cwa,Pu:2014fva}.
The transport coefficients for the CME and CESE $\xi$ and $\xi_{5}$
are \citep{Fukushima:2008xe,Gao:2012ix,Chen:2012ca}, 
\begin{equation}
\xi=eC\mu_{5},\;\xi_{5}=eC\mu_{e},\label{eq:CME_coefficient_01}
\end{equation}
where $\mu_{e}$ and $\mu_{5}$ are the chemical potentials for the
charge and chiral charge, respectively. The terms $\xi_{\omega}\omega^{\mu}$
and $\xi_{5\omega}\omega^{\mu}$ denote the chiral vortical effect
and the chiral current induced by the vortical fields \citep{Fukushima:2008xe,Gao:2012ix,Chen:2012ca},
with $\omega^{\mu}=\frac{1}{2}\epsilon^{\mu\nu\alpha\beta}u_{\nu}\partial_{\alpha}u_{\beta},$
being the vortical field.

Two types of Maxwell's equations are
\begin{eqnarray}
\partial_{\mu}F^{\mu\nu} & = & j_{e}^{\nu},\nonumber \\
\partial_{\mu}(\epsilon^{\mu\nu\alpha\beta}F_{\alpha\beta}) & = & 0.\label{eq:Maxwell}
\end{eqnarray}
Or, we can rewrite Eqs. (\ref{eq:Maxwell}) in the terms of $E^{\mu}$
and $B^{\mu}$, 
\begin{eqnarray}
\partial_{\mu}(E^{\mu}u^{\nu}-E^{\nu}u^{\mu}+\epsilon^{\mu\nu\alpha\beta}u_{\alpha}B_{\beta}) & = & j_{e}^{\nu},\nonumber \\
\partial_{\mu}(B^{\mu}u^{\nu}-B^{\nu}u^{\mu}+\epsilon^{\mu\nu\alpha\beta}u_{\beta}E_{\alpha}) & = & 0.\label{eq:Maxwell_02}
\end{eqnarray}

The thermodynamic relations and Gibbs relations read,
\begin{eqnarray}
\varepsilon+p & = & Ts+\mu n,\nonumber \\
d\varepsilon & = & Tds+\mu dn.\label{eq:thermo_01}
\end{eqnarray}
Note that, in general, the EM fields can modify the above relations
through the magnetization and electric polarization \citep{Pu:2016ayh}.
Here, we neglect these higher order corrections. Besides Eqs. (\ref{eq:thermo_01}),
we also need the equations of state (EoS) to close the whole system.

In Ref. \citep{Siddique:2019gqh}, the electric conductivity $\sigma$
is assumed as a constant. However, in general, the electric conductivity
depends on both temperature and chemical potential and is changed
with the evolution of the system. Therefore, in the current study,
we consider a temperature and/or energy density dependent conductivity
$\sigma$.

\section{equations in a Bjorken flow \label{sec:equations}}

In this section, we introduce two equations of state (EoS) and consider
the initial system is in a Bjorken flow. 

We follow Refs. \citep{Pu:2016ayh,Roy:2015kma,Pu:2016bxy,Pu:2016rdq,Siddique:2019gqh,Wang:2020qpx}
to search the solutions in a force-free MHD. We need to simplify the
main equations for MHD. First, we neglect the standard dissipative
terms in $T^{\mu\nu}$, i.e. we set $\Pi=\pi^{\mu\nu}=0$. Second,
although the vortical fields $\omega^{\mu}$ is of importance to the
spin polarization in the relativistic heavy ion collisions (see the
recent reviews \citep{Gao:2020vbh,Hidaka:2022dmn} and the references
therein), it is not directly connected to the evolution of EM fields.
Therefore, we also neglect the terms proportional to $\omega^{\mu}$
in the currents (\ref{eq:current_02}) for simplicity. Third, from
the Maxwell's equations, we know the charged fluid cells will be accelerated
by the EM fields. To search for the force-free type solutions in Ref.
\citep{Pu:2016ayh,Roy:2015kma,Pu:2016bxy,Pu:2016rdq,Siddique:2019gqh,Wang:2020qpx},
we assume that the fluid is charge neutral, i.e. we set $\mu_{e}=n_{e}=0$.
Since the chiral electric conductivity $\sigma_{5}$ is proportional
to $\mu_{e}\mu_{5}$ \citep{Huang:2013iia,Pu:2014cwa,Pu:2014fva}
in the small chemical potentials limits, we find that $\sigma_{5}=0$
under the assumption of vanishing $\mu_{e}$. 

We consider two typical EoS to close the system. 

In the high temperature limit, the EoS labeled as ``EOS-HT'' is
assumed as
\begin{eqnarray}
\varepsilon & = & c_{s}^{-2}p,\nonumber \\
n_{5} & = & a\mu_{5}T^{2},\label{eq:EoS_HT}
\end{eqnarray}
where speed of sound $c_{s}$ and $a$ are dimensionless constants.
For an ideal fluid, $c_{s}^{2}=a=1/3$ for instance \citep{Pu:2011vr}.
In the high temperature limit, the electric conductivity is assumed
to be proportional to temperature $T$ due to the dimension analysis,
i.e.
\begin{equation}
\sigma=\sigma_{0}\left[\frac{T(\tau)}{T(\tau_{0})}\right]+\mathcal{O}\left(\frac{\mu_{5}}{T}\right),\label{eq:Sigma_HT}
\end{equation}
with $\sigma_{0}$ being a constant, $\tau$ is the proper time and
$\tau_{0}$ being the initial proper time. 

In the high chiral chemical potential limit, we label the EoS as ``EOS-HC'',
which is given by,
\begin{eqnarray}
\varepsilon & = & c_{s}^{-2}p,\nonumber \\
n_{5} & = & a\mu_{5}^{3}.\label{eq:EoS_HC}
\end{eqnarray}
In the ideal fluid limit, $a=1/(3\pi^{2})$ \citep{Pu:2011vr,Gao:2012ix}.
The electric conductivity is assumed to be, $\sigma\propto(\mu_{e}^{2}+\mu_{5}^{2})/T$
, i.e.
\begin{equation}
\sigma=\sigma_{0}\left[\frac{n_{5}(\tau)}{n_{5}(\tau_{0})}\right]^{2/3}\left[\frac{\varepsilon(\tau_{0})}{\varepsilon(\tau)}\right]^{c_{s}^{2}/(1+c_{s}^{2})}+\mathcal{O}(T/\mu,T/\mu_{5}).\label{eq:Sigma_HC}
\end{equation}
Note that, for simplicity, we replace the temperature dependence by
the energy density dependent in Eq.~(\ref{eq:Sigma_HC}). More systematical
discussions on the electric conductivity can be found in lattice QCD
\citep{Aarts:2007wj,Ding:2010ga}, perturbative QCD at finite temperature
and chemical potentials and holographic models \citep{Chen:2013tra}
and other models \citep{Tuchin:2013ie}.


We summarize the main differential equations. The equations (\ref{eq:EMT-01},
\ref{eq:conservation}, \ref{eq:Maxwell}) with the constitution equations
(\ref{eq:Fuild_Tensor}, \ref{eq:current_02}), the EoS (\ref{eq:EoS_HT})
or (\ref{eq:EoS_HC}) with thermodynamic relations (\ref{eq:thermo_01})
and electric conductivity (\ref{eq:Sigma_HT}) or (\ref{eq:Sigma_HC})
are the main equations for the anomalous MHD. The acceleration equations
for the fluid velocity, i.e. $\Delta_{\mu\alpha}\partial_{\nu}T^{\mu\nu}=0$,
gives,
\begin{eqnarray}
(u\cdot\partial)u_{\alpha} & = & \frac{1}{\varepsilon+p+E^{2}+B^{2}}\left[\Delta_{\alpha}^{\ \nu}\partial_{\nu}(p+\frac{1}{2}E^{2}+\frac{1}{2}B^{2})\right.\nonumber \\
 &  & +\Delta_{\mu\alpha}(E\cdot\partial)E^{\mu}+E_{\alpha}(\partial\cdot E)+\Delta_{\mu\alpha}(B\cdot\partial)B^{\mu}\nonumber \\
 &  & +B_{\alpha}(\partial\cdot B)+\epsilon^{\nu\lambda\rho\sigma}E_{\lambda}B_{\rho}u_{\sigma}(\partial_{\nu}u_{\alpha})\nonumber \\
 &  & \left.+(\partial\cdot u)\epsilon_{\alpha\lambda\rho\sigma}E^{\lambda}B^{\rho}u^{\sigma}+\Delta_{\mu\nu}(u\cdot\partial)\epsilon^{\mu\lambda\rho\sigma}E_{\lambda}B_{\rho}u_{\sigma}\right].\label{eq:acceleration_eq_01}
\end{eqnarray}
The total energy conservation equation $u_{\mu}\partial_{\nu}T^{\mu\nu}=0$
reads, 
\begin{eqnarray}
 &  & (u\cdot\partial)\left(\varepsilon+\frac{1}{2}E^{2}+\frac{1}{2}B^{2}\right)+(\varepsilon+p+E^{2}+B^{2})(\partial\cdot u)\nonumber \\
 & = & u_{\mu}(E\cdot\partial)E^{\mu}+u_{\mu}(B\cdot\partial)B^{\mu}+\epsilon^{\nu\lambda\rho\sigma}\partial_{\nu}(E_{\lambda}B_{\rho}u_{\sigma})+u_{\mu}(\partial\cdot u)\epsilon^{\mu\lambda\rho\sigma}E_{\lambda}B_{\rho}u_{\sigma}.\label{eq:energy_con_01}
\end{eqnarray}
The conservation equations (\ref{eq:conservation}) becomes, 
\begin{eqnarray}
\partial_{\mu}(\sigma E^{\mu}+\xi B^{\mu}) & = & 0,\nonumber \\
\partial_{\mu}(n_{5}u^{\mu}) & = & 0.
\end{eqnarray}
The explicit expression for the Maxwell's equations (\ref{eq:Maxwell_02})
under our assumptions is, 
\begin{eqnarray}
\partial_{\mu}(E^{\mu}u^{\nu}-E^{\nu}u^{\mu}+\epsilon^{\mu\nu\alpha\beta}u_{\alpha}B_{\beta}) & = & \sigma E^{\nu}+\xi B^{\nu},\nonumber \\
\partial_{\mu}(B^{\mu}u^{\nu}-B^{\nu}u^{\mu}+\epsilon^{\mu\nu\alpha\beta}u_{\beta}E_{\alpha}) & = & 0.\label{eq:Maxwell_03}
\end{eqnarray}


We follow the basic idea in Refs. \citep{Pu:2016ayh,Roy:2015kma,Pu:2016bxy,Pu:2016rdq,Siddique:2019gqh,Wang:2020qpx}
to search for the analytic solutions in our cases. We assume that
the fluid is a Bjorken flow initially. We introduce the Milne coordinates
\begin{equation}
z=\tau\sinh\eta,\quad t=\tau\cosh\eta,
\end{equation}
where
\begin{eqnarray}
\tau & = & \sqrt{t^{2}-z^{2}},\nonumber \\
\eta & = & \frac{1}{2}\ln\left(\frac{t+z}{t-z}\right),
\end{eqnarray}
are the proper time and the space-time rapidity, respectively. The
fluid velocity of a longitudinal boost invariant Bjorken flow is given
by \citep{Bjorken:1982qr},
\begin{equation}
u^{\mu}=(\cosh\eta,0,0,\sinh\eta)=\gamma(1,0,0,z/t).\label{eq:Bjokren_velocity_01}
\end{equation}
For simplicity, we also assume the initial $E^{\mu}$ and $B^{\mu}$
depends on $\tau$ only and are parallel or anti-parallel to the each
other. Without loss of generality, we set the initial EM field be
put in the $y$ direction only,
\begin{equation}
E^{\mu}=(0,0,\chi E(\tau),0),\;B^{\mu}=(0,0,B(\tau),0),\label{eq:EB_02}
\end{equation}
where $\chi=\pm1$ represent parallel or anti-parallel. Following
the standard conclusion in a Bjorken flow, we also assume that all
the thermodynamic quantities depend on proper time $\tau$ only at
the initial time $\tau_{0}$.

We first assume that during the evolution the EM fields are still
satisfying the profile (\ref{eq:EB_02}) and the thermodynamic variables
still depend on the proper time only and then check whether this assumption
can always be satisfied. The acceleration equation (\ref{eq:acceleration_eq_01})
reduces to $(u\cdot\partial)u_{\alpha}=0$, i.e. the fluid will not
be accelerated by the EM fields. The Maxwell's equations (\ref{eq:Maxwell_03})
reduce to, 
\begin{eqnarray}
(u\cdot\partial)E+E(\partial\cdot u)+\sigma E+\chi\xi B & = & 0,\nonumber \\
(u\cdot\partial)B+B(\partial\cdot u) & = & 0.\label{eq:Max_02}
\end{eqnarray}
The equations for energy conservation and chiral anomaly becomes,
under these assumptions and with the help of Eq. (\ref{eq:Max_02}),
\begin{eqnarray}
(u\cdot\partial)\varepsilon+(\varepsilon+p)(\partial\cdot u)-\sigma E^{2}-\chi\xi EB & = & 0,\nonumber \\
(u\cdot\partial)n_{5}+n_{5}(\partial\cdot u)-e^{2}C\chi EB & = & 0.\label{eq:Con_Eqs}
\end{eqnarray}
The charge conservation equation $\partial_{\mu}j^{\mu}=\partial_{\mu}(\sigma E^{\mu}+\xi B^{\mu})=0$
will automatically be satisfied under the assumptions. Combining Eqs.
(\ref{eq:Max_02}) and (\ref{eq:Con_Eqs}), we can conclude that with
the initial configurations (\ref{eq:EB_02}), the initial fluid velocity
(\ref{eq:Bjokren_velocity_01}) holds and all the thermodynamic variables
will always depend on the proper time only during the evolution.

Before end of this section, we would like to emphasis the $E^{\mu}$
and $B^{\mu}$ in Eq. (\ref{eq:EB_02}) are defined in a comoving
frame. The electromagnetic fields \textbf{$\mathbf{E}_{\textrm{lab}}$
}and \textbf{$\mathbf{B}_{\textrm{lab}}$} in the laboratory frame
can be obtained through $F^{\mu\nu}$ directly, 
\begin{eqnarray}
\mathbf{E}_{\textrm{lab}} & = & (\gamma v^{z}B(\tau),\chi\gamma E(\tau),0),\nonumber \\
\mathbf{B}_{\textrm{lab}} & = & (-\gamma v^{z}\chi E(\tau),\gamma B(\tau),0).
\end{eqnarray}

In next section, we will solve Eqs. (\ref{eq:Max_02}, \ref{eq:Con_Eqs}).
Interestingly, there are no terms proportional to $\partial_{\mu}\sigma$
in Eqs. (\ref{eq:Max_02}, \ref{eq:Con_Eqs}), i.e. Eqs. (\ref{eq:Max_02},
\ref{eq:Con_Eqs}) are the same as those in the case of constant $\sigma$.
We emphasis that now the $\sigma$ depends on proper time, e.g. in
Eq. (\ref{eq:Sigma_HT}) for EoS-HT (\ref{eq:EoS_HT}) and Eq. (\ref{eq:Sigma_HC})
for EoS-HC (\ref{eq:EoS_HC}). Therefore, the system evolution becomes
more complicated then those in Ref. \citep{Siddique:2019gqh,Wang:2020qpx}. 


\section{Solutions in high temperature limit \label{sec:solution_HT}}

In this section, we solve the simplified differential equations (\ref{eq:Max_02},
\ref{eq:Con_Eqs}) with the EoS-HT (\ref{eq:EoS_HT}) and temperature
dependent conductivity (\ref{eq:Sigma_HT}). 

Following Ref. \citep{Csorgo:2003rt,Shokri:2017xxn,Siddique:2019gqh,Wang:2020qpx},
we implement the following method. For a given differential equation
\begin{equation}
\left(\frac{d}{d\tau}+\frac{m}{\tau}\right)f(\tau)=f(\tau)\frac{d}{d\tau}\lambda(\tau),\label{eq:f(x)_01-1}
\end{equation}
with $m$ being a constant and $\lambda(\tau)$ being a given source
term, the solution for $f(\tau)$ can be written in a compact form,
\begin{equation}
f(\tau)=f(\tau_{0})\exp[\lambda(\tau)-\lambda(\tau)]\left(\frac{\tau_{0}}{\tau}\right)^{m},\label{eq:f(x)_02}
\end{equation}
where $\tau_{0}$ is an initial proper time and $f(\tau_{0})$ is
the initial value of $f(\tau)$ at $\tau_{0}$.

From Eq. (\ref{eq:Max_02}) it is straightforward to get 
\begin{equation}
B(\tau)=B_{0}\frac{\tau_{0}}{\tau},\label{eq:B_field}
\end{equation}
where $B_{0}$ is initial magnetic fields. 

We rewrite Eqs (\ref{eq:Max_02}, \ref{eq:Con_Eqs}) in a compact
form, 
\begin{eqnarray}
\frac{d}{d\tau}\varepsilon+(1+c_{s}^{2})\frac{\varepsilon}{\tau} & = & \varepsilon\frac{d}{d\tau}\mathcal{L},\nonumber \\
\frac{d}{d\tau}E+\frac{E}{\tau} & = & E\frac{d}{d\tau}\mathcal{E},\nonumber \\
\frac{d}{d\tau}n_{5}+\frac{n_{5}}{\tau} & = & n_{5}\frac{d}{d\tau}\mathcal{N},\label{eq:EN_01}
\end{eqnarray}
where the source terms are given by,
\begin{eqnarray}
\frac{d}{d\tau}\mathcal{L} & = & \frac{1}{\varepsilon}\sigma E^{2}+\frac{1}{\varepsilon}eC\chi\mu_{5}EB,\nonumber \\
\frac{d}{d\tau}\mathcal{E} & = & -\sigma-eC\chi\mu_{5}\frac{B}{E},\nonumber \\
\frac{d}{d\tau}\mathcal{N} & = & \frac{e^{2}C\chi EB}{n_{5}}.\label{eq:EN_02}
\end{eqnarray}
Following Eq. (\ref{eq:f(x)_02}), the solutions of Eqs. (\ref{eq:EN_01})
read, 
\begin{eqnarray}
E(\tau) & = & E_{0}\left(\frac{\tau_{0}}{\tau}\right)\exp\left[\mathcal{E}(\tau)-\mathcal{E}(\tau_{0})\right]\equiv E_{0}\left(\frac{\tau_{0}}{\tau}\right)x(\tau),\nonumber \\
n_{5}(\tau) & = & n_{5,0}\left(\frac{\tau_{0}}{\tau}\right)\exp\left[\mathcal{N}(\tau)-\mathcal{N}(\tau_{0})\right]\equiv n_{5,0}\left(\frac{\tau_{0}}{\tau}\right)y(\tau),\nonumber \\
\varepsilon(\tau) & = & \varepsilon_{0}\left(\frac{\tau_{0}}{\tau}\right)^{1+c_{s}^{2}}\exp\left[\mathcal{L}(\tau)-\mathcal{L}(\tau_{0})\right]\equiv\varepsilon_{0}\left(\frac{\tau_{0}}{\tau}\right)^{1+c_{s}^{2}}z(\tau),\label{eq:EN_03}
\end{eqnarray}
 where $E_{0},n_{5,0},\varepsilon_{0}$ are initial electric field,
chiral density and energy density at $\tau=\tau_{0}$ and we have
introduced three variables $x(\tau)$, $y(\tau)$ and $z(\tau)$,
\begin{eqnarray}
x(\tau) & = & \exp\left[\mathcal{E}(\tau)-\mathcal{E}(\tau_{0})\right],\nonumber \\
y(\tau) & = & \exp\left[\mathcal{N}(\tau)-\mathcal{N}(\tau_{0})\right],\nonumber \\
z(\tau) & = & \exp\left[\mathcal{L}(\tau)-\mathcal{L}(\tau_{0})\right].\label{eq:xyz_01}
\end{eqnarray}

In the high temperature limit, the energy density can be written in
the power series of $\mu_{5}/T\ll1$, 
\begin{equation}
\varepsilon\left(\tau\right)=\varepsilon_{0}\left(\frac{T}{T_{0}}\right)^{1+c_{s}^{-2}}+\mathcal{O}\left(\frac{\mu_{5}^{2}}{T^{2}}\right).\label{eq:energy_01}
\end{equation}
From Eq. (\ref{eq:EN_03}) for energy density $\varepsilon$, we get
\begin{equation}
T=T_{0}\left(\frac{\tau_{0}}{\tau}\right)^{c_{s}^{2}}z(\tau)^{c_{s}^{2}/(1+c_{s}^{2})}.\label{eq:T-tau relation}
\end{equation}
and, from Eq. (\ref{eq:Sigma_HT}), 
\begin{equation}
\sigma=\sigma_{0}\left(\frac{\tau_{0}}{\tau}\right)^{c_{s}^{2}}z(\tau)^{c_{s}^{2}/(1+c_{s}^{2})}.\label{eq:Sigma_02}
\end{equation}

Inserting the expression for $x,y,z$ into Eqs. (\ref{eq:EN_02}),
yields, 
\begin{eqnarray}
\frac{d}{d\tau}x & = & -\sigma_{0}\left(\frac{\tau_{0}}{\tau}\right)^{c_{s}^{2}}xz{}^{c_{s}^{2}/(1+c_{s}^{2})}-\frac{a_{1}}{\tau_{0}}\left(\frac{\tau_{0}}{\tau}\right)^{1-2c_{s}^{2}}yz^{-2c_{s}^{2}/(1+c_{s}^{2})},\nonumber \\
\frac{d}{d\tau}y & = & a_{2}\frac{x}{\tau},\nonumber \\
\frac{d}{d\tau}z & = & \frac{E_{0}^{2}}{\varepsilon_{0}}\sigma_{0}\left(\frac{\tau_{0}}{\tau}\right)x^{2}z^{c_{s}^{2}/(1+c_{s}^{2})}+\frac{a_{1}}{\tau_{0}}\left(\frac{E_{0}^{2}}{\varepsilon_{0}}\right)\left(\frac{\tau_{0}}{\tau}\right)^{2-3c_{s}^{2}}xyz^{-2c_{s}^{2}/(1+c_{s}^{2})},\label{eq:Final_01}
\end{eqnarray}
where by definition, $x(\tau_{0})=y(\tau_{0})=z(\tau_{0})=1$, and
$a_{1,2}$ are all dimensionless constants determined by the initial
conditions 
\begin{eqnarray}
a_{1} & = & eC\chi\frac{B_{0}n_{5,0}}{aT_{0}^{2}E_{0}}\tau_{0},\nonumber \\
a_{2} & = & e^{2}C\chi\frac{E_{0}B_{0}}{n_{5,0}}\tau_{0}.\label{eq:a123_01}
\end{eqnarray}
We notice that $a_{1,2}\propto C\sim\hbar$ and it denotes that the
terms proportional to $a_{1},a_{2}$ in Eqs. (\ref{eq:Final_01})
are quantum corrections to the ordinary MHD. 

\begin{figure}
\includegraphics[scale=0.25]{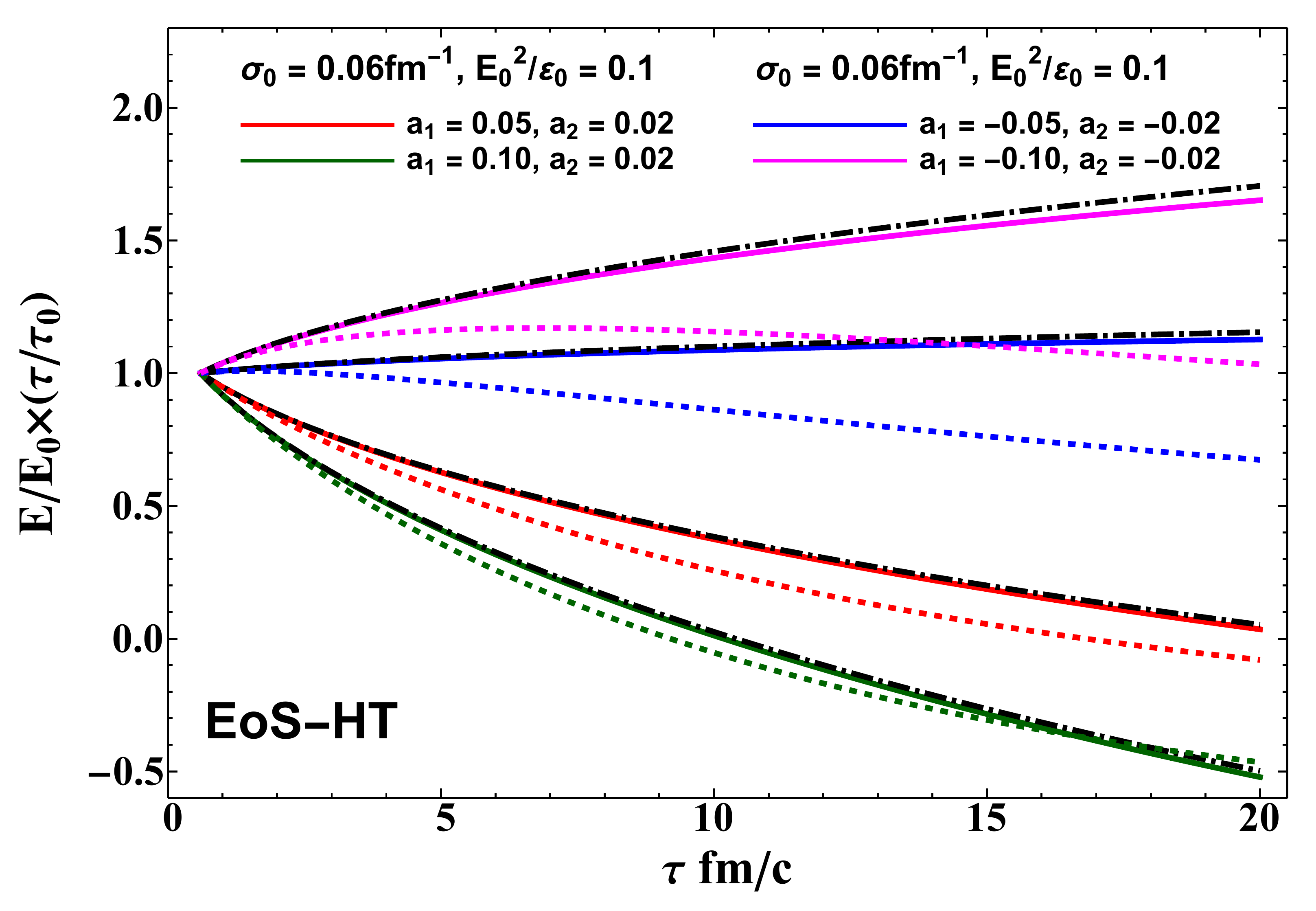}\includegraphics[scale=0.25]{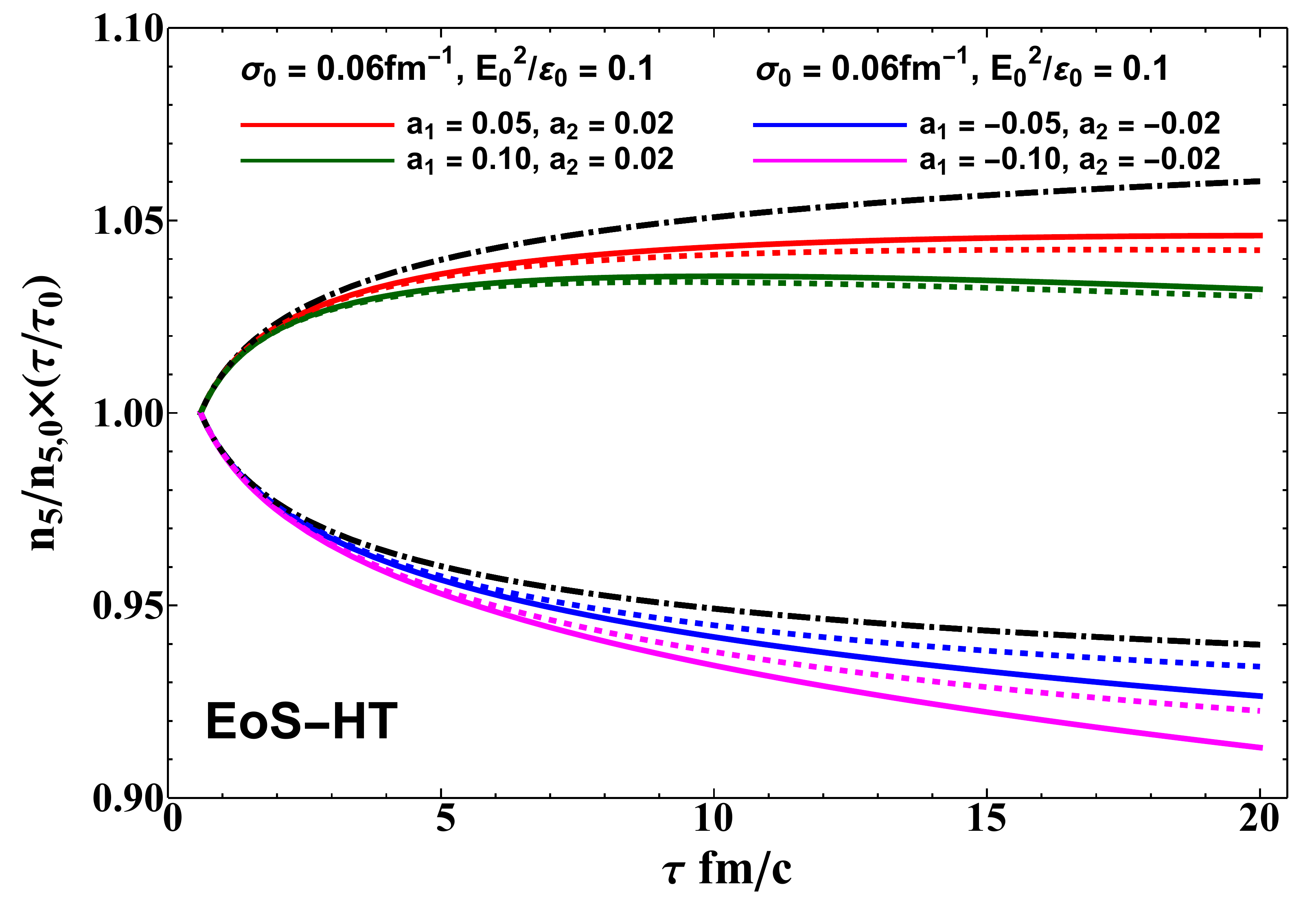}

\includegraphics[scale=0.25]{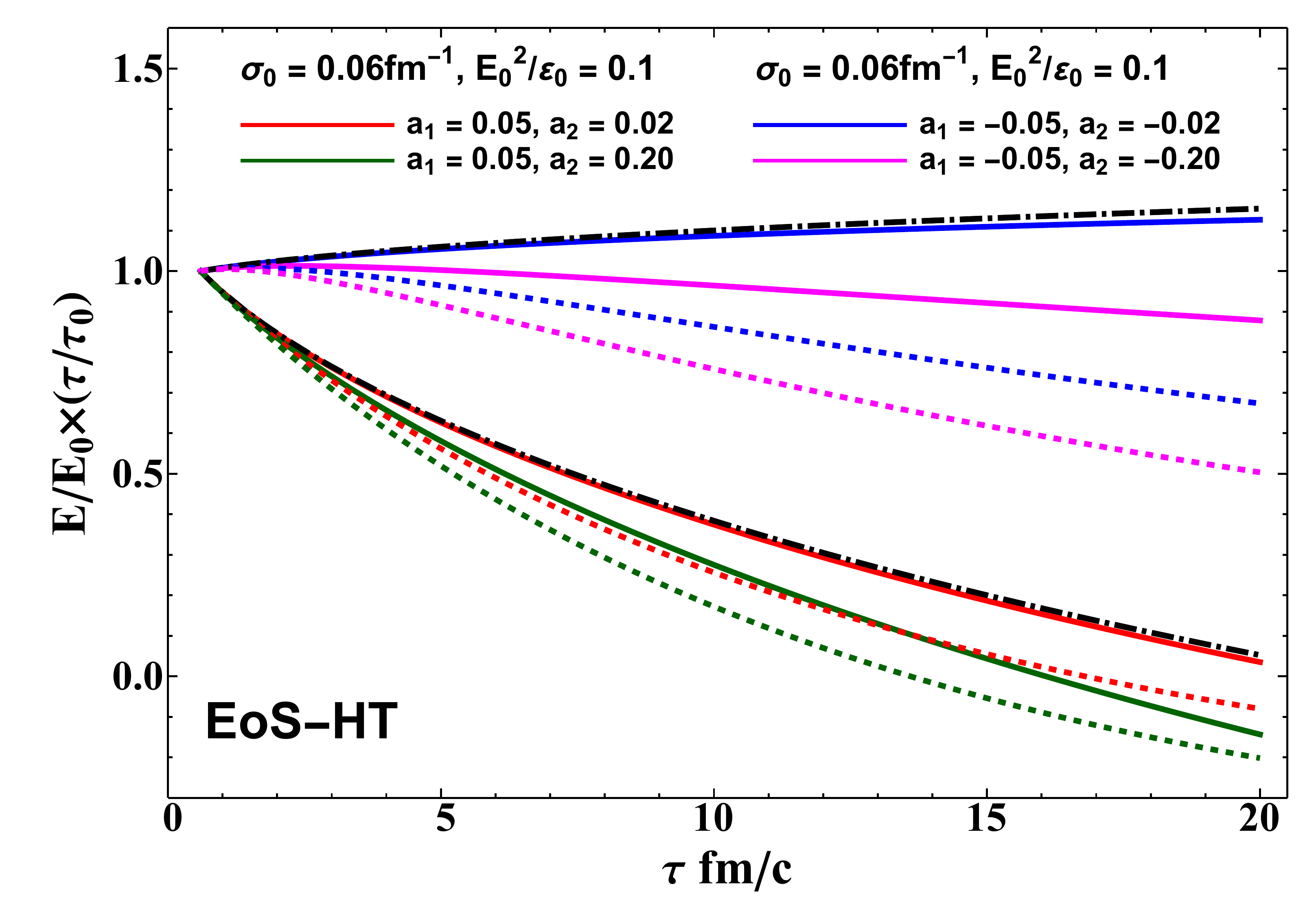}\includegraphics[scale=0.25]{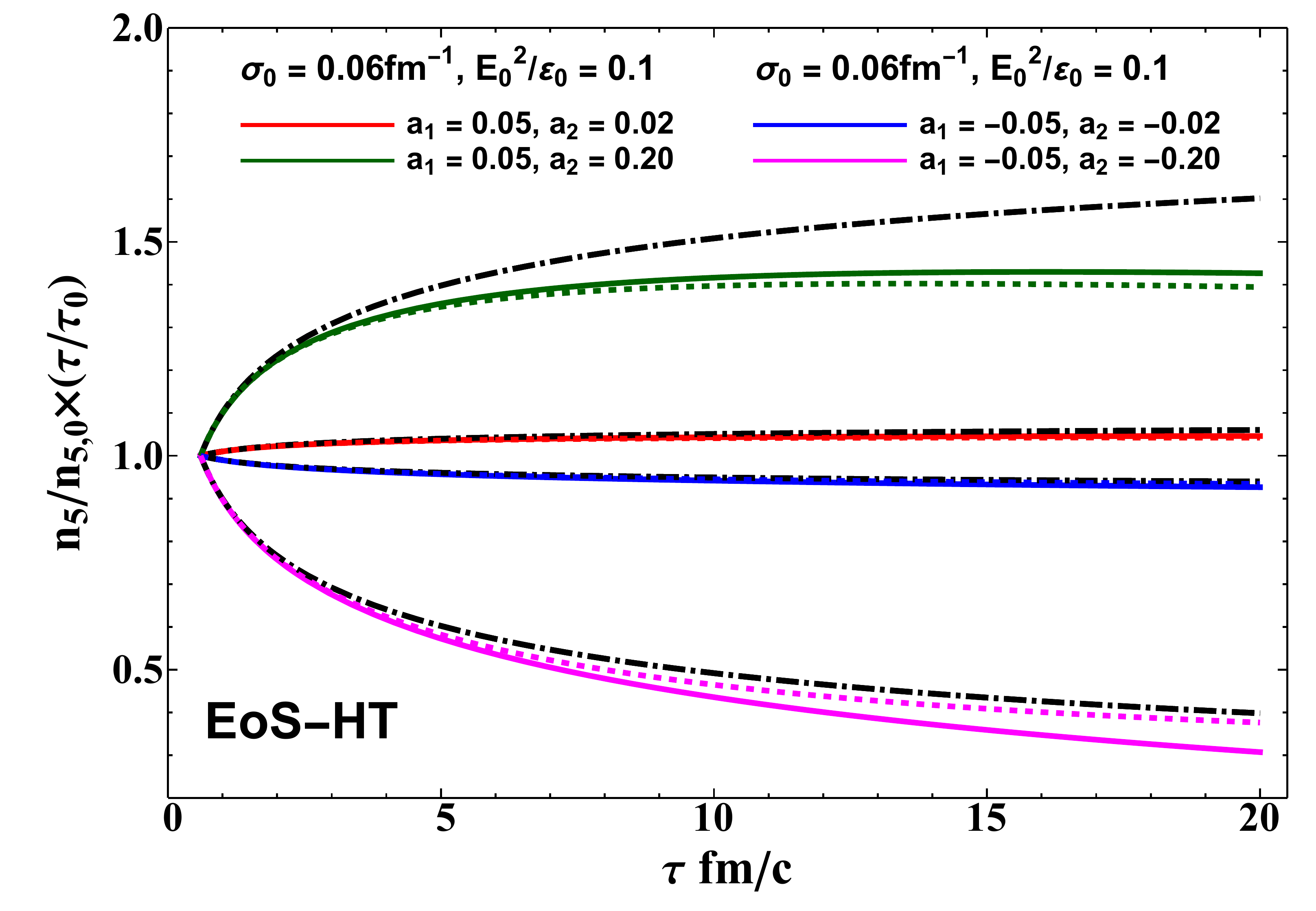}

\includegraphics[scale=0.25]{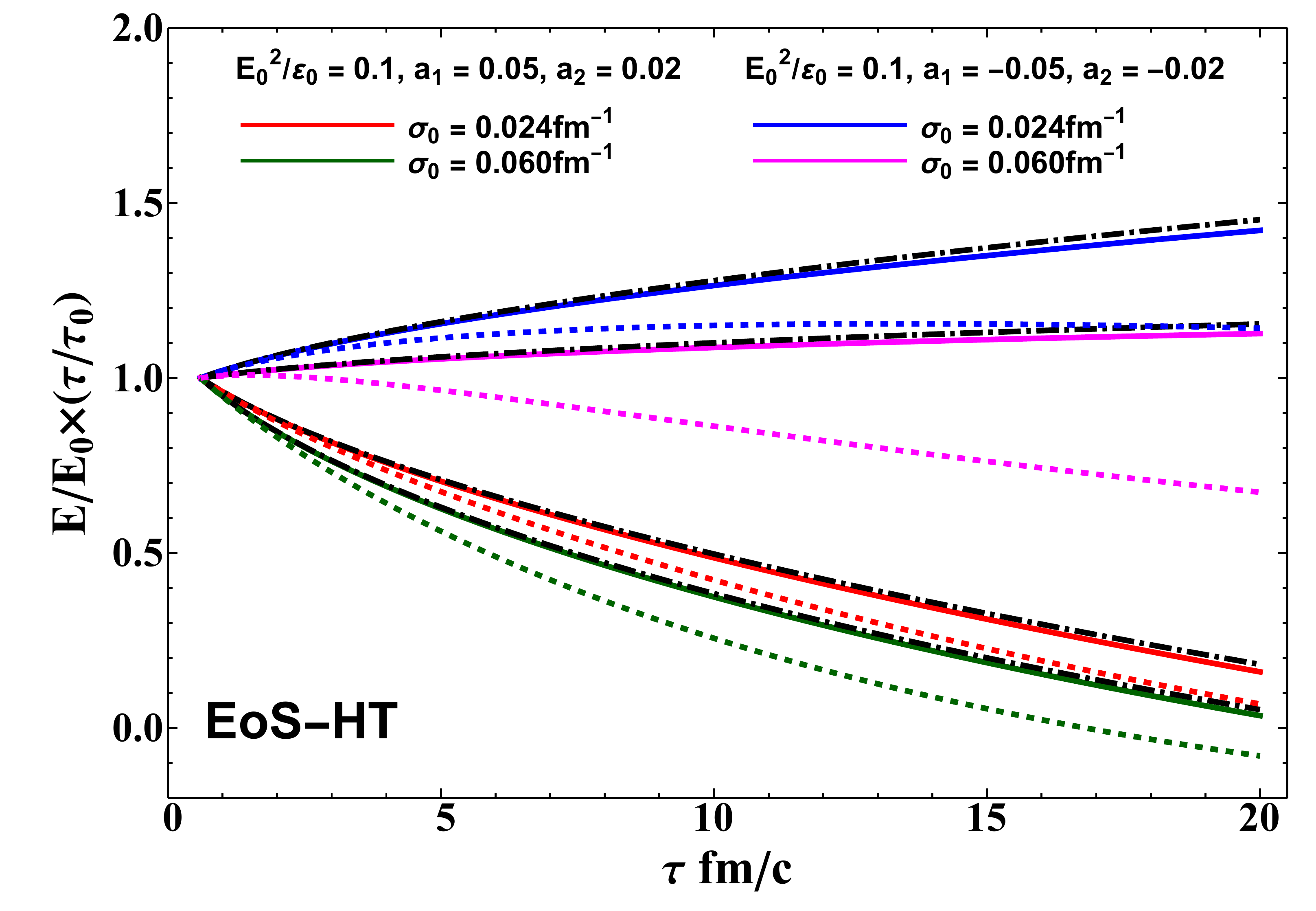}\includegraphics[scale=0.25]{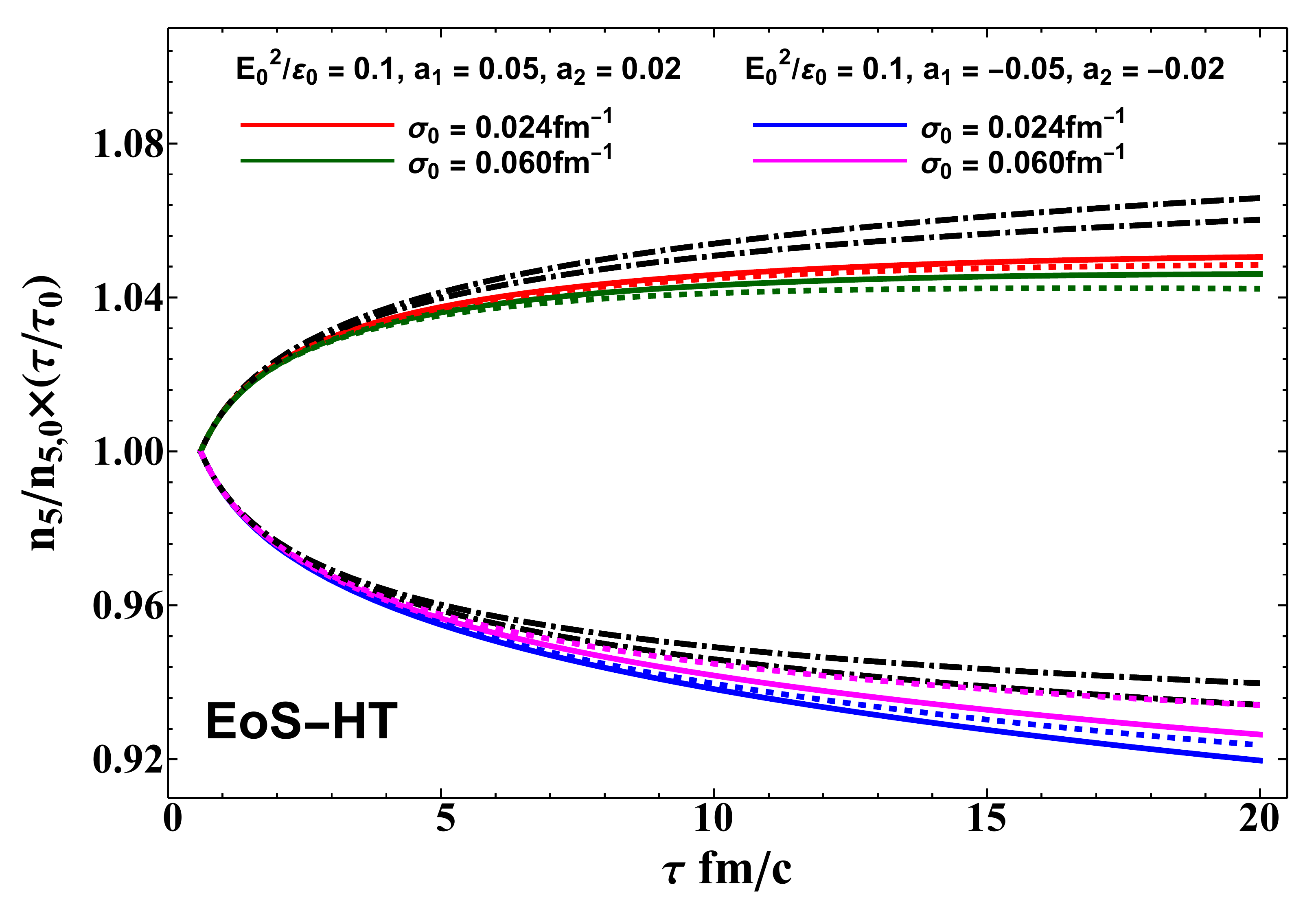}

\caption{The proper time scaled electric field $E/E_{0}\times\left(\tau/\tau_{0}\right)$
and chiral charge density $n_{5}/n_{5,0}\times\left(\tau/\tau_{0}\right)$
computed with EoS-HC (\ref{eq:EoS_HT}) as functions of proper time
$\tau$ with different parameters $a_{1},a_{2},\sigma_{0}$. We have
fixed $E_{0}^{2}/\varepsilon_{0}=0.1$. The colored solid, black dashed-dotted
and colored dotted lines denote the results for solving the results
from Eqs. (\ref{eq:EN_01}, \ref{eq:EN_02}) numerically, analytic
solutions (\ref{eq:EN_03}, \ref{eq:Soln_xyz_01}) and numerical results
with a constant $\sigma$ derived in Ref. \citep{Siddique:2019gqh,Wang:2020qpx},
respectively. \label{fig:EM_T}}
\end{figure}

\begin{figure}
\includegraphics[scale=0.25]{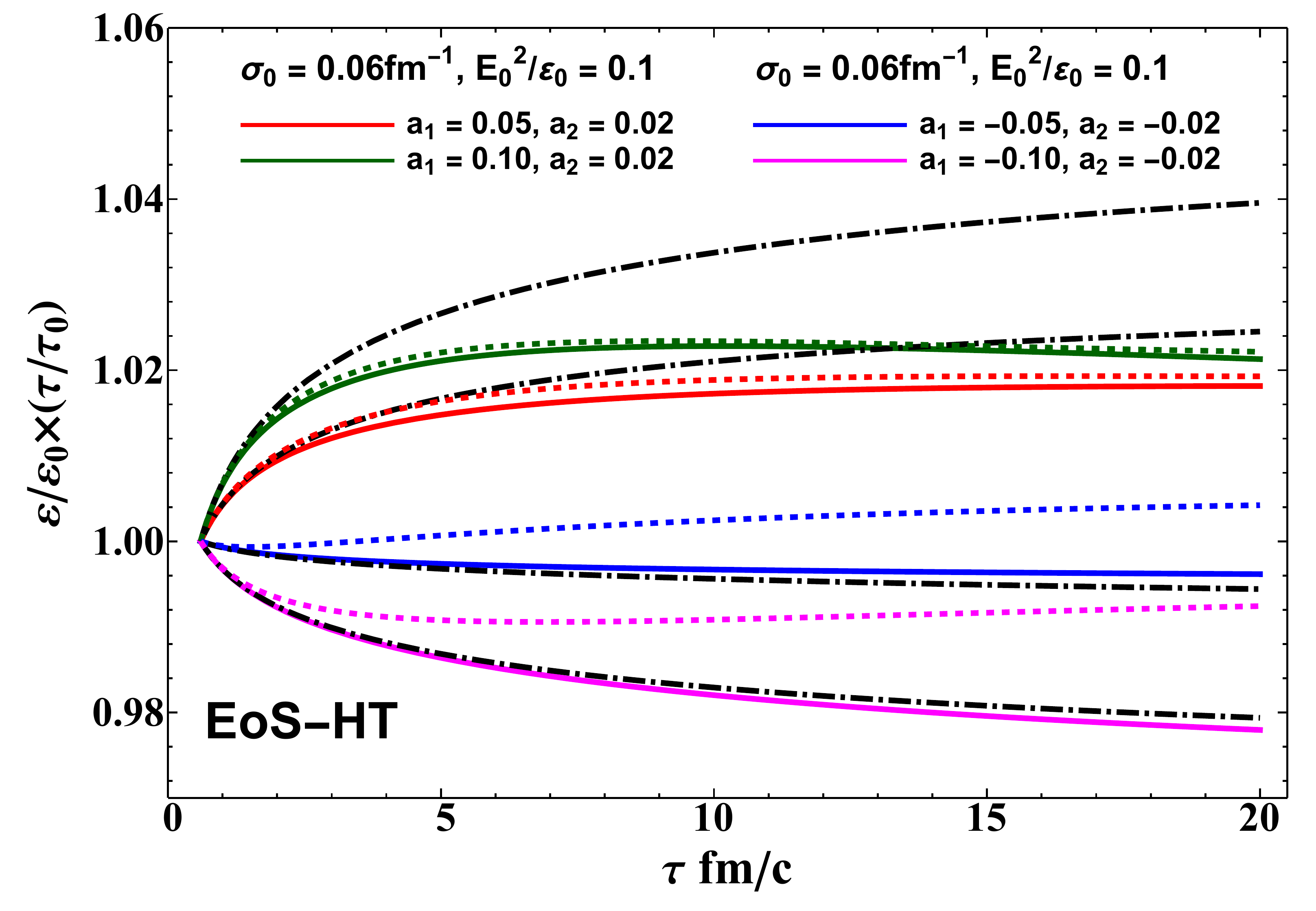}\includegraphics[scale=0.25]{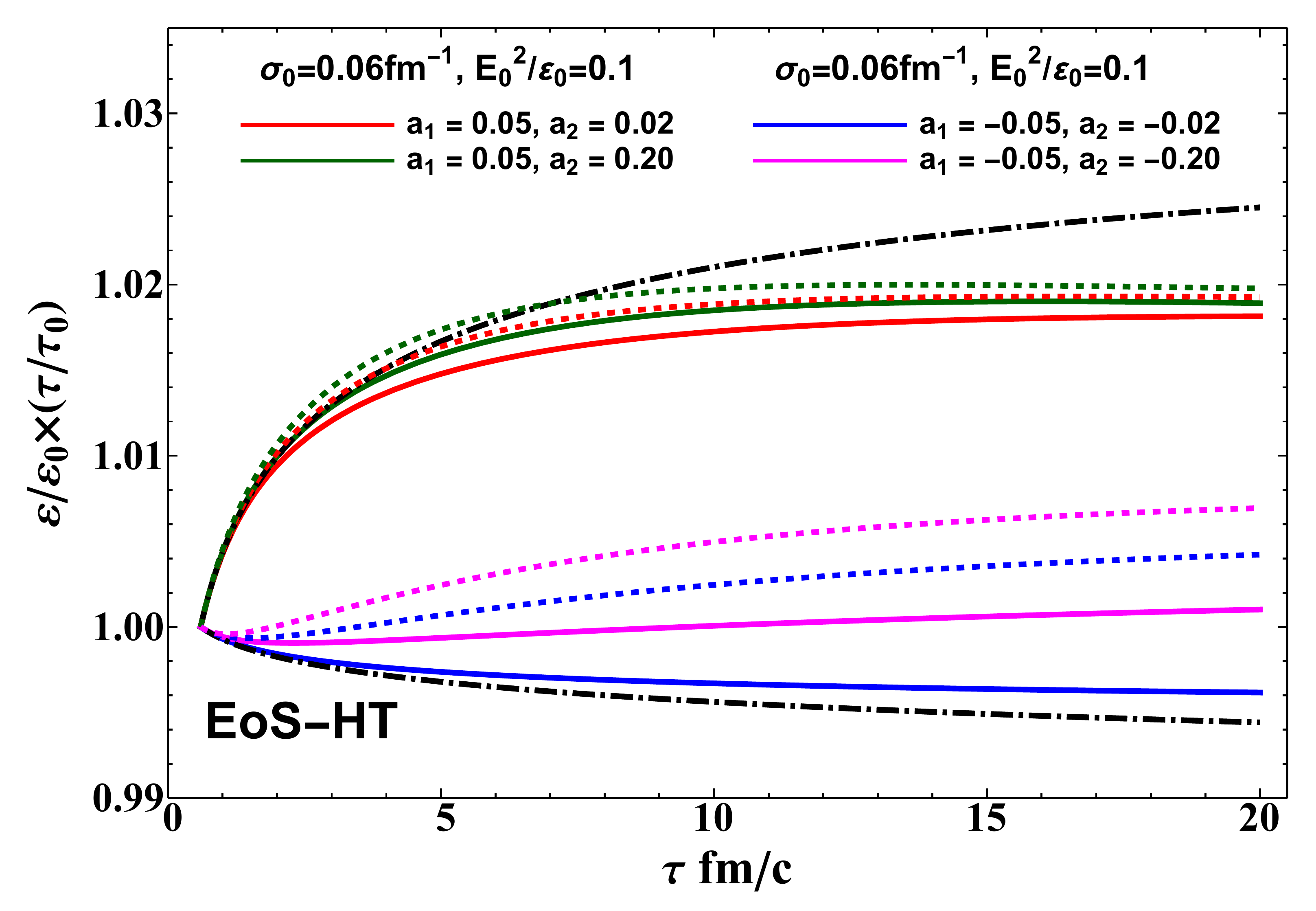}

\includegraphics[scale=0.25]{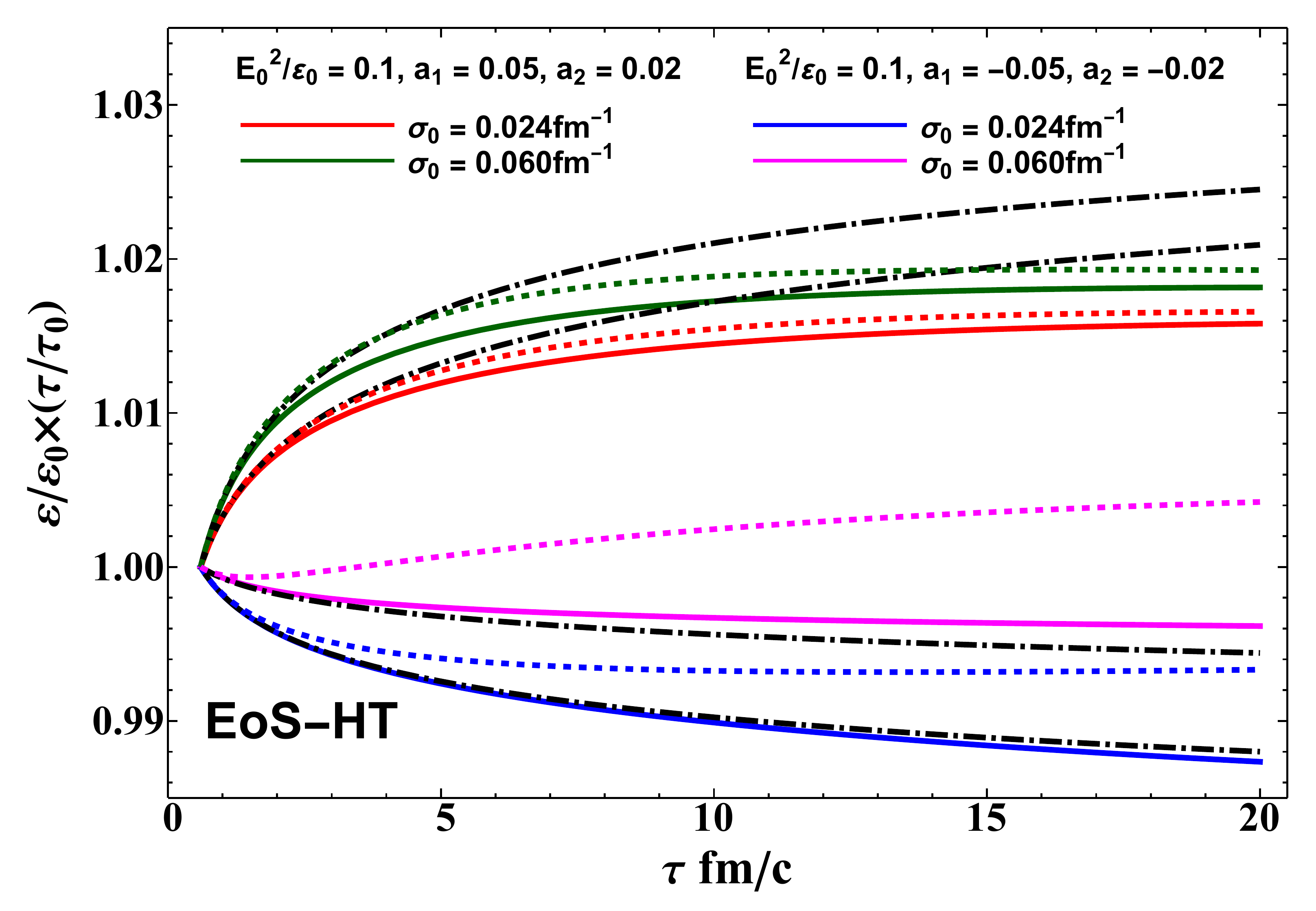}\includegraphics[scale=0.25]{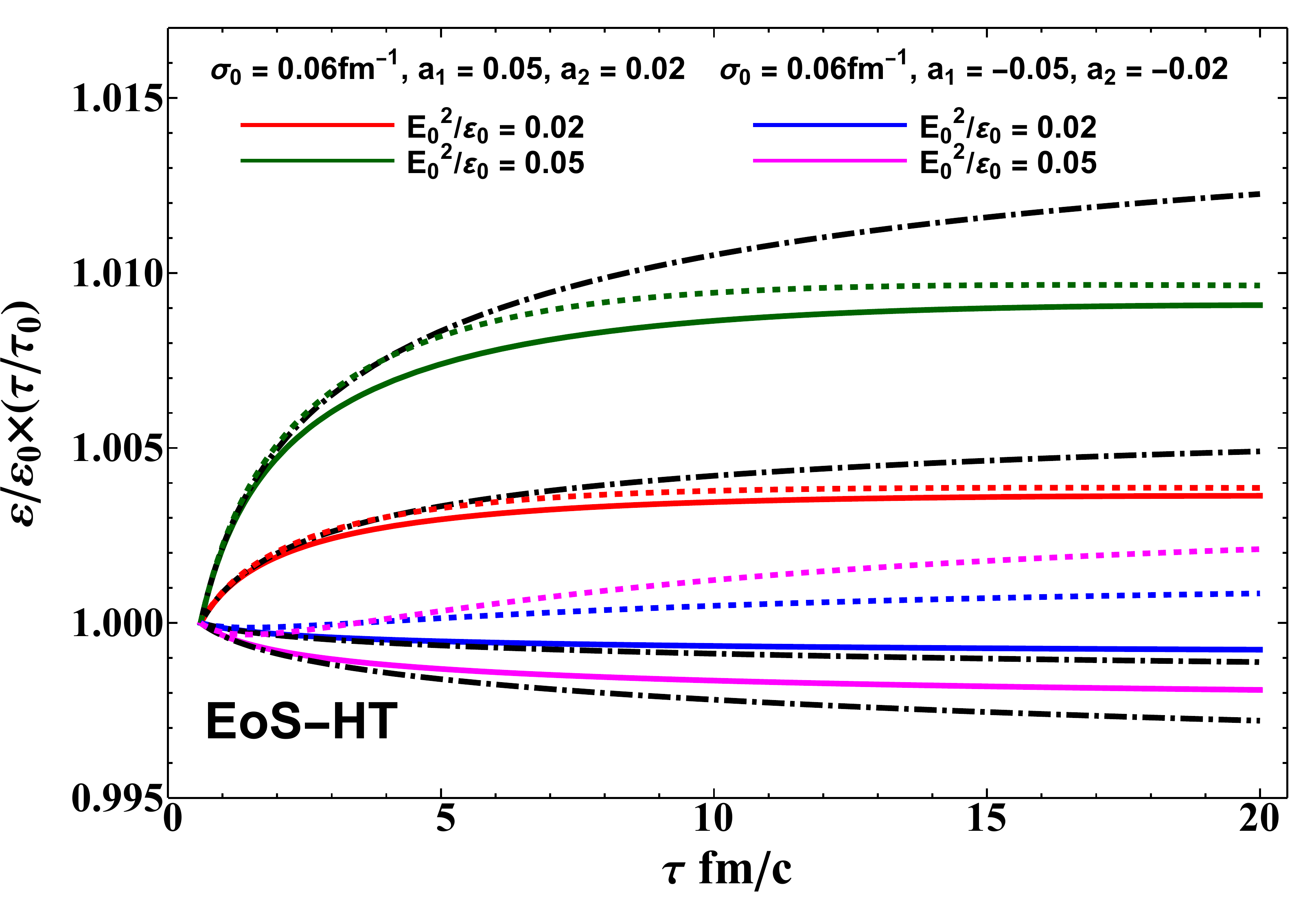}

\caption{The proper time scaled energy density $\varepsilon/\varepsilon_{0}\times\left(\tau/\tau_{0}\right)^{1+c_{s}^{2}}$
computed with EoS-HT (\ref{eq:EoS_HT}) as functions of proper time
$\tau$ with different parameters $a_{1},a_{2},\sigma_{0}$ and $E_{0}^{2}/\varepsilon_{0}$.
The colored solid, black dashed-dotted and colored dotted lines denote
the results for solving the results from Eqs. (\ref{eq:EN_01}, \ref{eq:EN_02})
numerically, analytic solutions (\ref{eq:EN_03}, \ref{eq:Soln_xyz_01})
and numerical results with a constant $\sigma$ derived in Ref. \citep{Siddique:2019gqh,Wang:2020qpx},
respectively. \label{fig:en_T}}
\end{figure}

From the numerical simulations in Ref. \citep{Roy:2015coa}, we find
that the initial energy density of fluid $\varepsilon_{0}$ is much
larger than the initial energy density of EM fields $\varepsilon_{0}\gg E_{0}^{2},B_{0}^{2},E_{0}B_{0}$.
For simplicity, we consider $\mathcal{O}\left(\frac{E_{0}^{2}}{\varepsilon_{0}}\right)\sim\mathcal{O}(a_{1,2})\sim\mathcal{O}(\hbar)$.
To the leading order of $\hbar$, we get the solutions for Eqs. (\ref{eq:Final_01}),
\begin{eqnarray}
x(\tau) & = & e^{m_{1}(\tilde{\tau}-1)}-a_{1}e^{m_{1}\tilde{\tau}}(1-c_{s}^{2})^{-1}[\mathcal{E}(n_{1},m_{1})-\tilde{\tau}^{-2c_{s}^{2}/(c_{s}^{2}-1)}\mathcal{E}(n_{1},m_{1}\tilde{\tau})],\nonumber \\
y(\tau) & = & 1+a_{2}e^{-m_{1}}(1-c_{s}^{2})^{-1}[\mathcal{E}(1,-m_{1})-\mathcal{E}(1,-m_{1}\tilde{\tau})],\nonumber \\
z(\tau) & = & 1-\frac{E_{0}^{2}}{\varepsilon_{0}}m_{1}e^{-2m_{1}}[\mathcal{E}(1,-2m_{1})-\mathcal{E}(1,-2m_{1}\tilde{\tau})]\nonumber \\
 &  & +a_{1}\frac{E_{0}^{2}}{\varepsilon_{0}}e^{-m_{1}}(1-c_{s}^{2})^{-1}[\mathcal{E}(2n_{2},-m_{1})-\tilde{\tau}^{(1-3c_{s}^{2})/(c_{s}^{2}-1)}\mathcal{E}(2n_{2},-m_{1}\tilde{\tau})],\label{eq:Soln_xyz_01}
\end{eqnarray}
where 
\begin{equation}
\mathcal{E}(n,m)\equiv\int_{1}^{\infty}dt\,t^{-n}e^{-mt},
\end{equation}
 is the generated exponential integral, and 
\begin{eqnarray}
n_{1} & = & \frac{1-3c_{s}^{2}}{1-c_{s}^{2}},\nonumber \\
n_{2} & = & \frac{1-2c_{s}^{2}}{1-c_{s}^{2}},\nonumber \\
m_{1} & = & -\frac{\sigma_{0}\tau_{0}}{1-c_{s}^{2}},\nonumber \\
\tilde{\tau} & = & \left(\frac{\tau_{0}}{\tau}\right)^{c_{s}^{2}-1}.
\end{eqnarray}
Inserting the solutions for $x,y,z$ back into Eqs. (\ref{eq:EN_03}),
we can get the solutions for $E(\tau),n_{5}(\tau)$ and $\varepsilon(\tau)$. 

In the order of $\hbar^{0}$, we find,
\begin{eqnarray}
E(\tau) & = & E_{0}\left(\frac{\tau_{0}}{\tau}\right)x(\tau)\propto\frac{1}{\tau}e^{-\sigma(\tau)\tau},\nonumber \\
n_{5}(\tau) & = & n_{5,0}\left(\frac{\tau_{0}}{\tau}\right)y(\tau)\propto\frac{\tau_{0}}{\tau},\nonumber \\
\varepsilon(\tau) & = & \varepsilon_{0}\left(\frac{\tau_{0}}{\tau}\right)^{1+c_{s}^{2}}z(\tau)\propto\left(\frac{\tau_{0}}{\tau}\right)^{1+c_{s}^{2}},\label{eq:LO_solution_HT}
\end{eqnarray}
which are consistent with the results in a ordinary MHD with finite
electric conductivity in a Bjorken flow as discussed by Ref. \citep{Siddique:2019gqh,Wang:2020qpx}.
We also find that, 
\begin{equation}
\sigma(\tau)=\sigma_{0}\left[\frac{T(\tau)}{T(\tau_{0})}\right]\propto\left(\frac{\tau_{0}}{\tau}\right)^{c_{s}^{2}},
\end{equation}
which decays with proper time. The numerical studies of $\sigma(\tau)$
in the evolution of QGP can also be found in Ref. \citep{Zhang:2022lje,Wang:2021oqq}.

Next, we present the above perturbative analytic solutions (\ref{eq:EN_03},
\ref{eq:Soln_xyz_01}) and the results for solving Eqs. (\ref{eq:EN_01})
directly in a numerical way. We choose the $\tau_{0}=0.6\textrm{ fm/c}$
and the speed of sound $c_{s}^{2}=1/3$. The electric conductivity
$\sigma$ are given by $\sigma\sim5.8T/T_{c}\,\textrm{MeV}$ from
lattice QCD \citep{Aarts:2007wj,Ding:2010ga,Tuchin:2013ie} and $\sigma\sim20-30\,\textrm{MeV}$
for $T\simeq200\,\textrm{MeV}$ from the holographic QCD \citep{Pu:2014cwa,Pu:2014fva}.
In this work, we choose the region of the initial electric conductivity
$\sigma_{0}$ as $\sigma_{0}\sim5-30\,\textrm{MeV}\simeq0.025-0.15\ \mathrm{fm}^{-1}$.

From the solutions (\ref{eq:Soln_xyz_01}), the proper time scaled
electric field $E/E_{0}\times\left(\tau/\tau_{0}\right)$ and chiral
density $n_{5}/n_{5,0}\times\left(\tau/\tau_{0}\right)$ will not
be sensitive to the $E_{0}^{2}/\varepsilon_{0}$. We have also confirmed
it numerically. For simplicity, we fix $E_{0}^{2}/\varepsilon_{0}=0.1$
for the discussion on $E/E_{0}\times\left(\tau/\tau_{0}\right)$ and
$n_{5}/n_{5,0}\times\left(\tau/\tau_{0}\right)$.

In left handed side of Fig. \ref{fig:EM_T}, we plot the proper time
scaled electric field $E/E_{0}\times\left(\tau/\tau_{0}\right)$ as
functions of the proper time $\tau$ with different sets of parameters
$a_{1,2},\sigma_{0}$. The analytic solutions (\ref{eq:EN_03}, \ref{eq:Soln_xyz_01})
agree with the numerical results well. Note that our analytic solutions
(\ref{eq:Soln_xyz_01}) for $E$ do not have the $a_{2}$ dependence
and, therefore, it only matches the numerical results in small $|a_{2}|$
case. 

Similar to Ref. \citep{Siddique:2019gqh,Wang:2020qpx}, we find that
$E(\tau)$ decays faster when $a_{1}$ increases and $E/E_{0}$ can
be negative at late time on condition that the initial $E$ and $B$
field are of the same orientation due to the competition between the
anomalous conservation equation and Maxwell's equations. 

We can also compare the our results (solid or dashed-dotted liens)
with temperature dependent $\sigma(\tau)$ to those with a constant
$\sigma$ derived in Ref. \citep{Siddique:2019gqh,Wang:2020qpx} (dotted
lines). From Eq. (\ref{eq:LO_solution_HT}) in the order of $\hbar^{0}$,
$E(\tau)\sim\frac{1}{\tau}e^{-\sigma_{0}\tau_{0}^{c_{s}^{2}}\tau^{1-c_{s}^{2}}}\sim\frac{1}{\tau}e^{-\sigma_{0}\tau_{0}^{1/3}\tau^{2/3}}\leq\frac{1}{\tau}e^{-\sigma_{0}\tau_{0}}$
when $\tau\geq\tau_{0}$. It means that the $\sigma(\tau)$ will quicken
up the decaying of $E(\tau)$. We can understand it in the following
way. At late time limit, $\lim_{\tau\rightarrow\infty}\sigma(\tau)\rightarrow0$,
the fluid becomes dilute close to the vacuum and the EM fields decays much rapidity
in the vacuum than in a medium. It is consistent with the results
in Fig. \ref{fig:EM_T}. On the other hand, we also notice that when
$a_{1,2}$ are negative, the $E/E_{0}\times\left(\tau/\tau_{0}\right)$
may increase when proper time grows. It corresponds to the cases that
EM fields gain the energy from the medium.


In right handed side of Fig. \ref{fig:EM_T}, we present the proper
time scaled chiral density $n_{5}/n_{5,0}\times\left(\tau/\tau_{0}\right)$
as functions of the proper time $\tau$ with different sets of parameters
$a_{1},a_{2}$ and $\sigma_{0}$. Since our analytic solutions (\ref{eq:Soln_xyz_01})
for $n_{5}$ do not have the $a_{1}$ dependence, the analytic solutions
for $n_{5}$ agree with the numerical results in small $|a_{1}|$
limit. We find that the proper time scaled chiral density is not sensitive
to the $a_{1}$ and $\sigma_{0}$ at early proper time. It agrees
with the analytic solutions (\ref{eq:Soln_xyz_01}). At late proper
time, proper time scaled chiral density decreases with $a_{1}$ decreasing.
We also observe that the proper time scaled chiral density increases
when $a_{2}$ grows. The temperature dependent $\sigma(\tau)$ seems
not to affect the evolution of $n_{5}$, which can also be found in
the solutions (\ref{eq:Soln_xyz_01}). 

In Figs. \ref{fig:en_T}, we plot the proper time scaled energy density
$\varepsilon/\varepsilon_{0}\times\left(\tau/\tau_{0}\right)^{1+c_{s}^{2}}$
as functions of the proper time $\tau$ with different sets of parameters
$a_{1},a_{2},\sigma_{0}$ and $E_{0}^{2}/\varepsilon_{0}$. 

As shown in Eqs. (\ref{eq:Soln_xyz_01}), our analytic solutions for
$\varepsilon$ do not have $a_{2}$ dependence and it only corresponds
to the small $|a_{2}|$ limit. On the other hand, we observe that
our analytic solutions (\ref{eq:Soln_xyz_01}) for $\varepsilon$
matches the numerical results only at early proper time. It implies
that the system is beyond the approximation for $\varepsilon$ in
Eq. (\ref{eq:energy_01}) at late proper time. 

We observe that the proper time scaled energy density increases when
$a_{1},a_{2},\sigma_{0},E_{0}^{2}/\varepsilon_{0}$ increases. When
$a_{1},a_{2},\sigma_{0},E_{0}^{2}/\varepsilon_{0}$ is large enough,
the proper time scaled energy density can increase with proper time.
It corresponds that the medium gains the energy from the EM fields,
which is also found in the ideal MHD with a background magnetic field
\citep{Roy:2015kma,Pu:2016ayh}. In the comparison with the case of
constant $\sigma_{0}$, we find that the temperature dependent $\sigma(\tau)$
enhances the decaying of energy density. 

In this section, we have solved the MHD with the temperature dependent
$\sigma(\tau$) in Eq. (\ref{eq:Sigma_HT}) and EoS-HT (\ref{eq:EoS_HT}).
We present both the analytic solutions (\ref{eq:Soln_xyz_01}) up
to the $\mathcal{O}(\hbar)$ and the numerical results. 

\section{Solutions in high chiral chemical potential limit \label{sec:solution_HC}}

In this section, we solve the simplified differential equations (\ref{eq:Max_02},
\ref{eq:Con_Eqs}) with the EoS-HC (\ref{eq:EoS_HC}) and temperature
and energy density dependent conductivity (\ref{eq:Sigma_HC}). 

\begin{figure}
\includegraphics[scale=0.25]{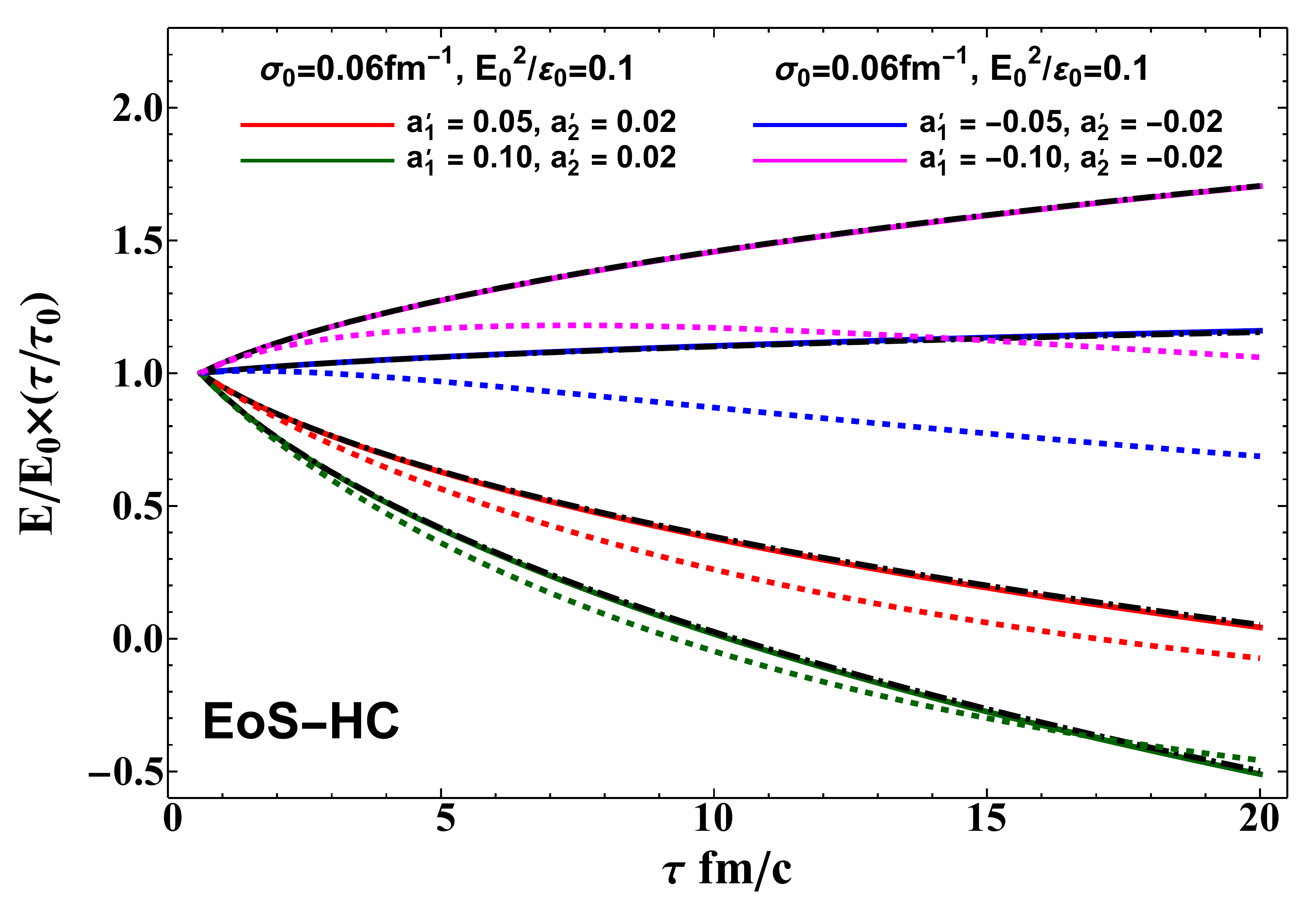}\includegraphics[scale=0.25]{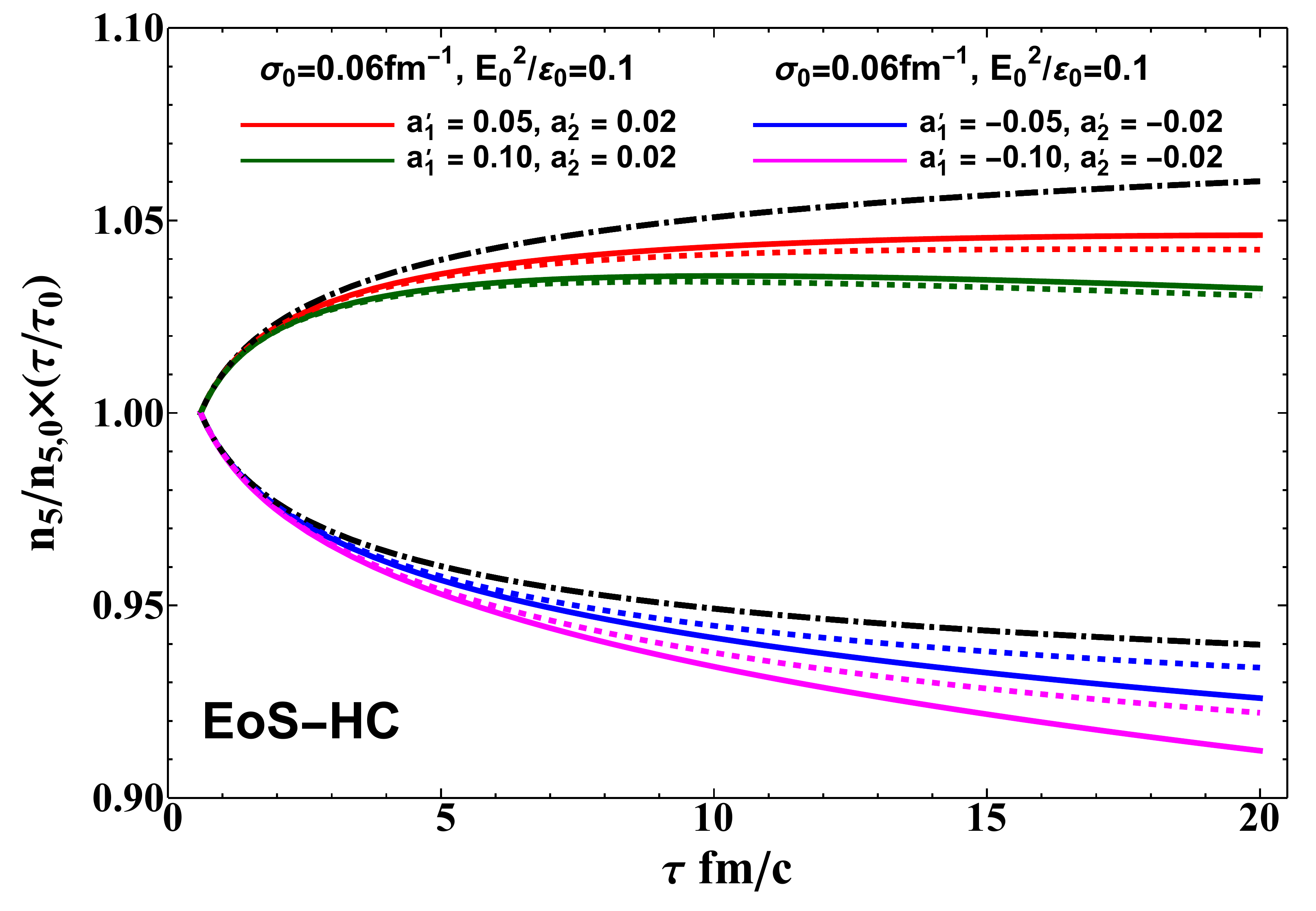}

\includegraphics[scale=0.25]{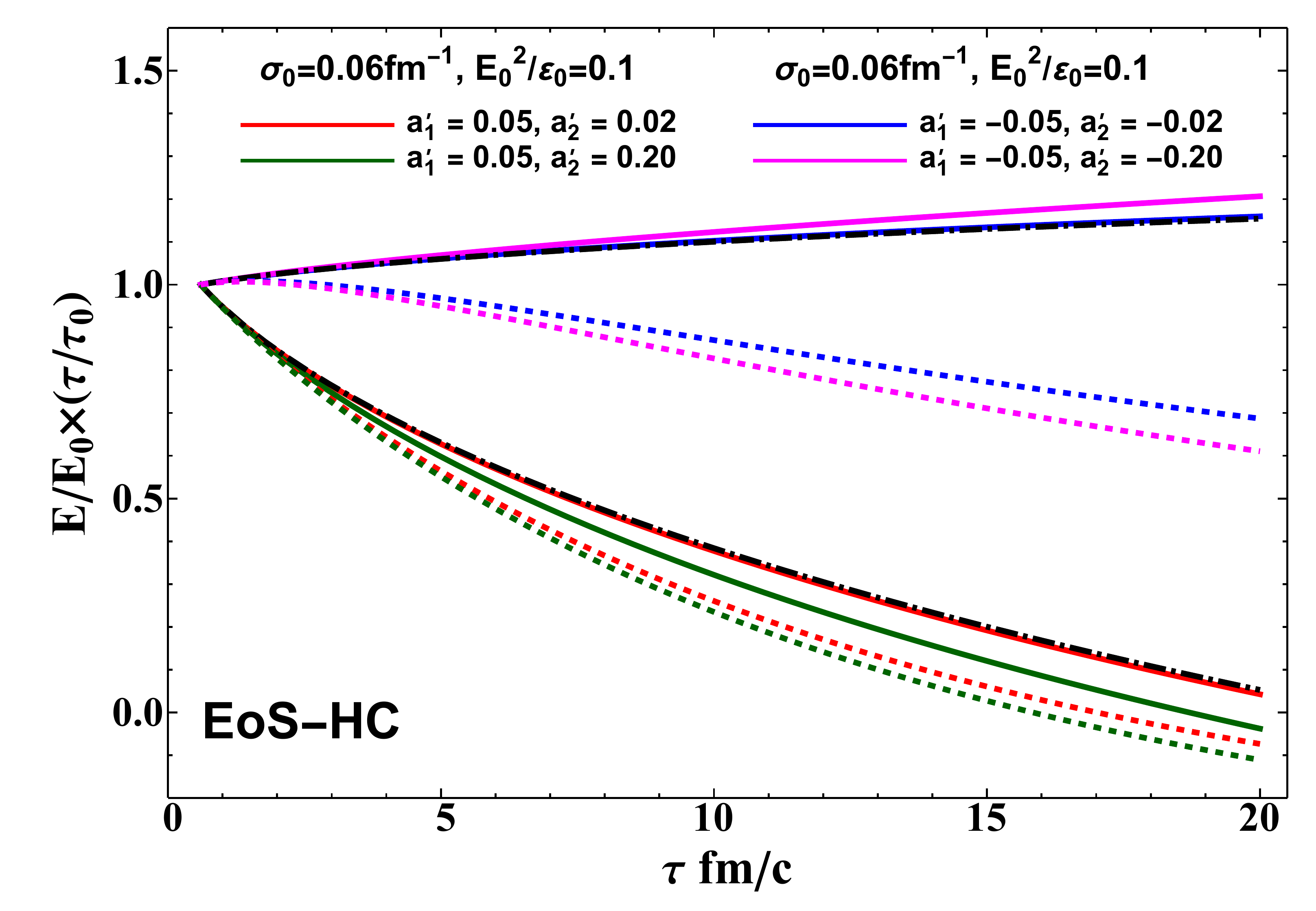}\includegraphics[scale=0.25]{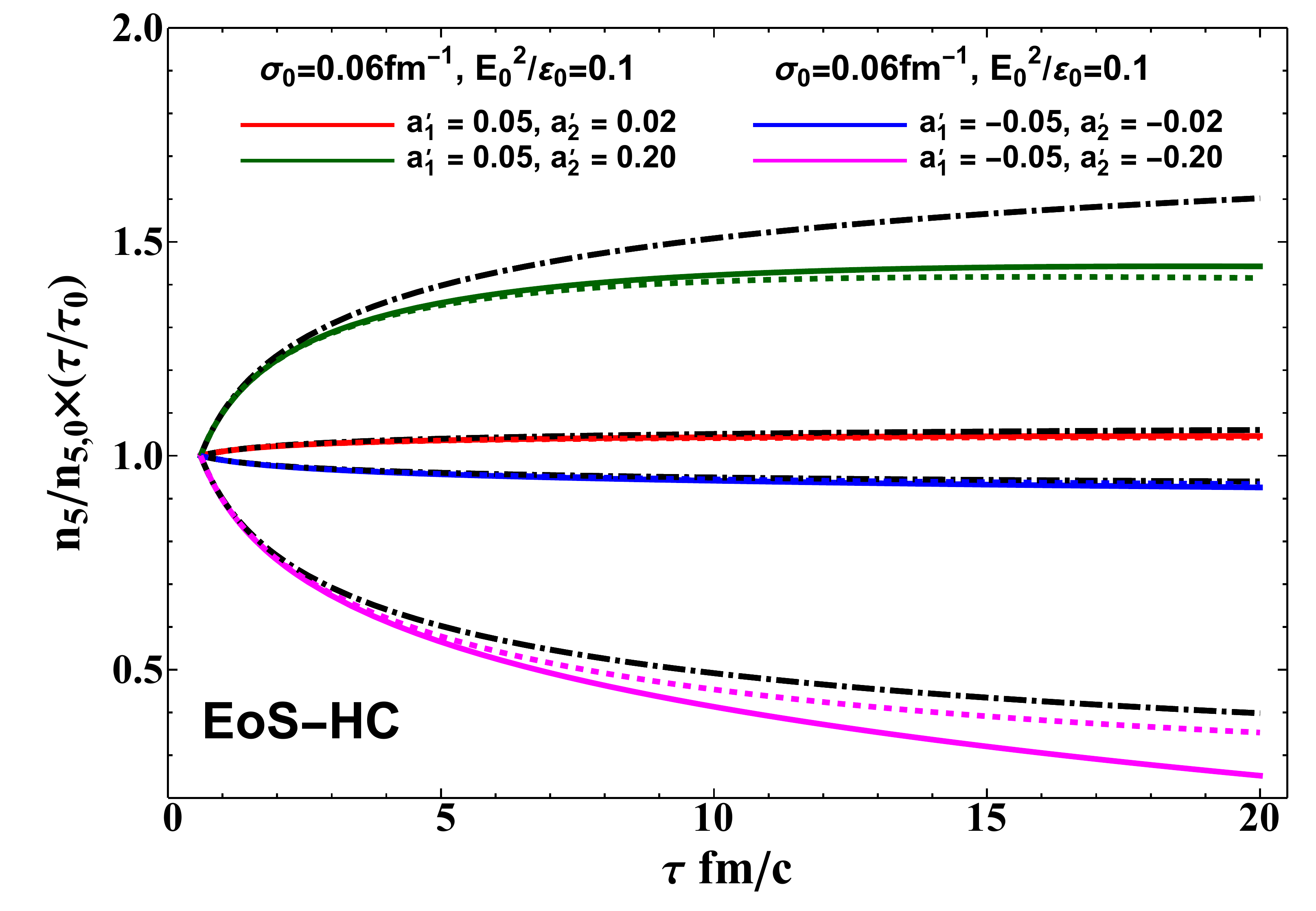}

\includegraphics[scale=0.25]{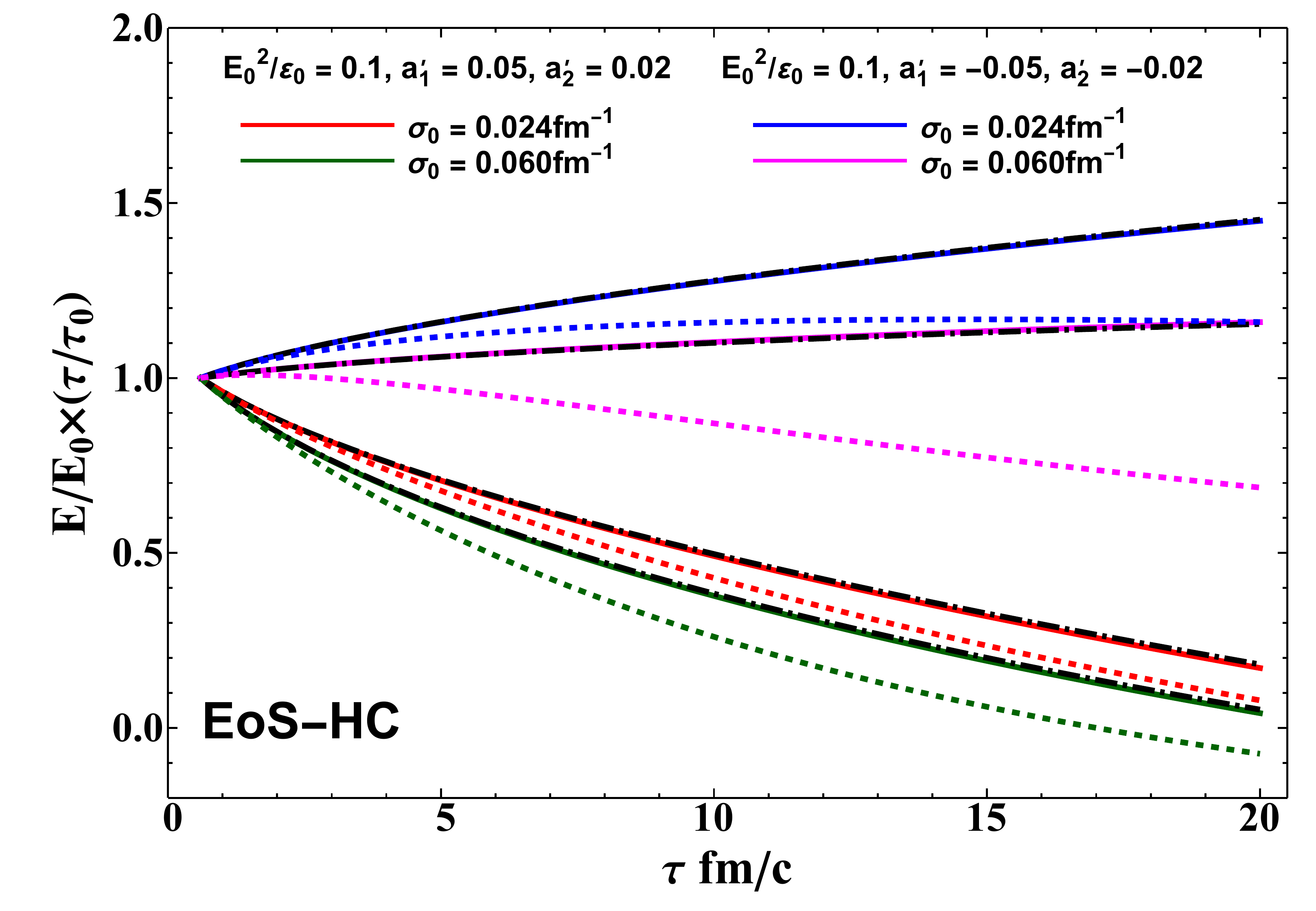}\includegraphics[scale=0.25]{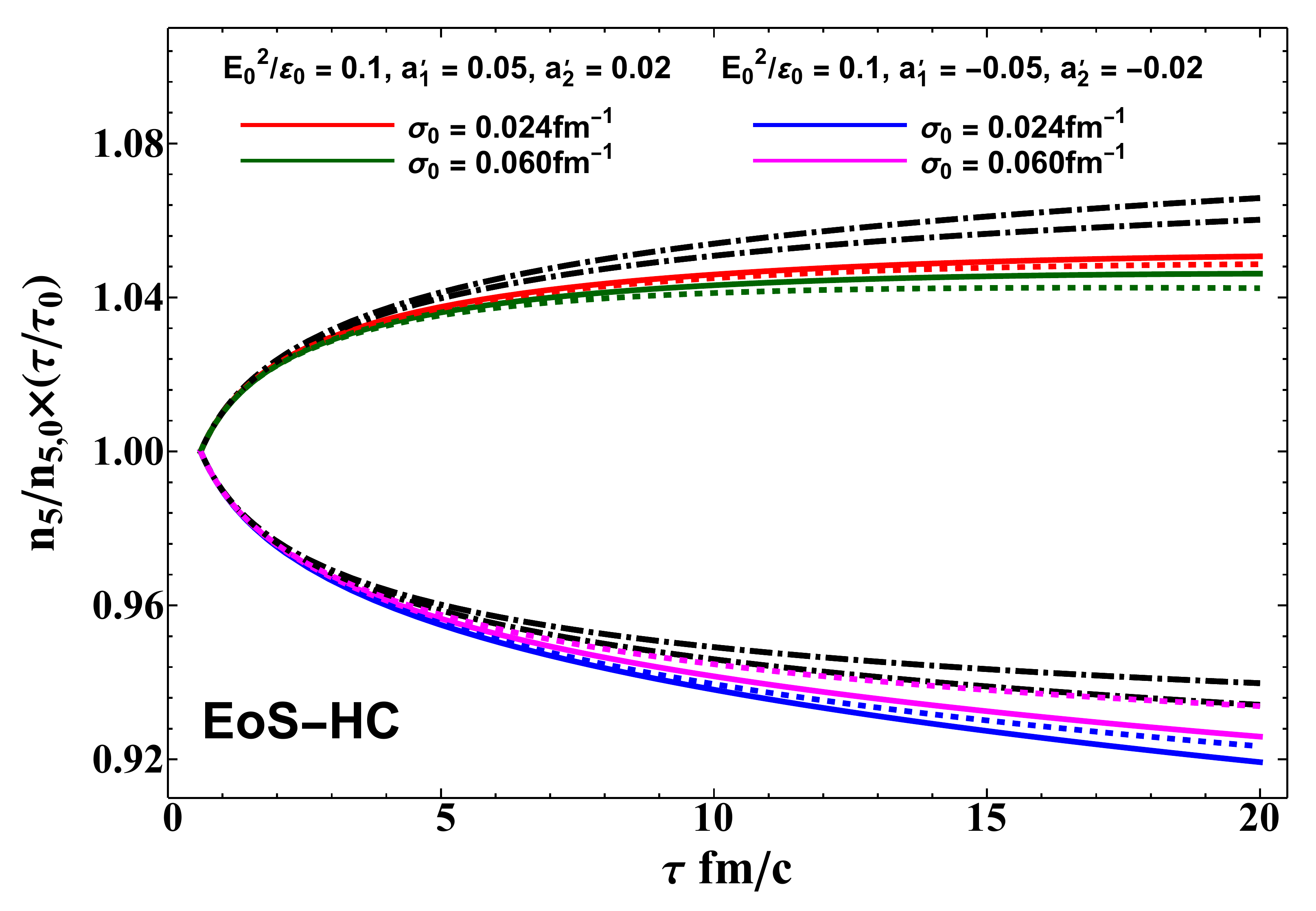}

\caption{The proper time scaled electric field $E/E_{0}\times\left(\tau/\tau_{0}\right)$
and chiral density $n_{5}/n_{5,0}\times\left(\tau/\tau_{0}\right)$
computed with EoS-HC (\ref{eq:EoS_HC}) as functions of proper time
$\tau$ with different parameters $a_{1},a_{2},\sigma_{0}$. For simplicity,
we fix $E_{0}^{2}/\varepsilon_{0}=0.1$. The colored solid, black
dashed-dotted and colored dotted lines denote the results for solving
the results from Eqs. (\ref{eq:EN_01}, \ref{eq:EN_02}) numerically,
analytic solutions (\ref{eq:EN_03}, \ref{eq:soln_xyz_02}) and numerical
results with a constant $\sigma$ derived in Ref. \citep{Siddique:2019gqh,Wang:2020qpx},
respectively. \label{fig:EM_p}}
\end{figure}

\begin{figure}
\includegraphics[scale=0.25]{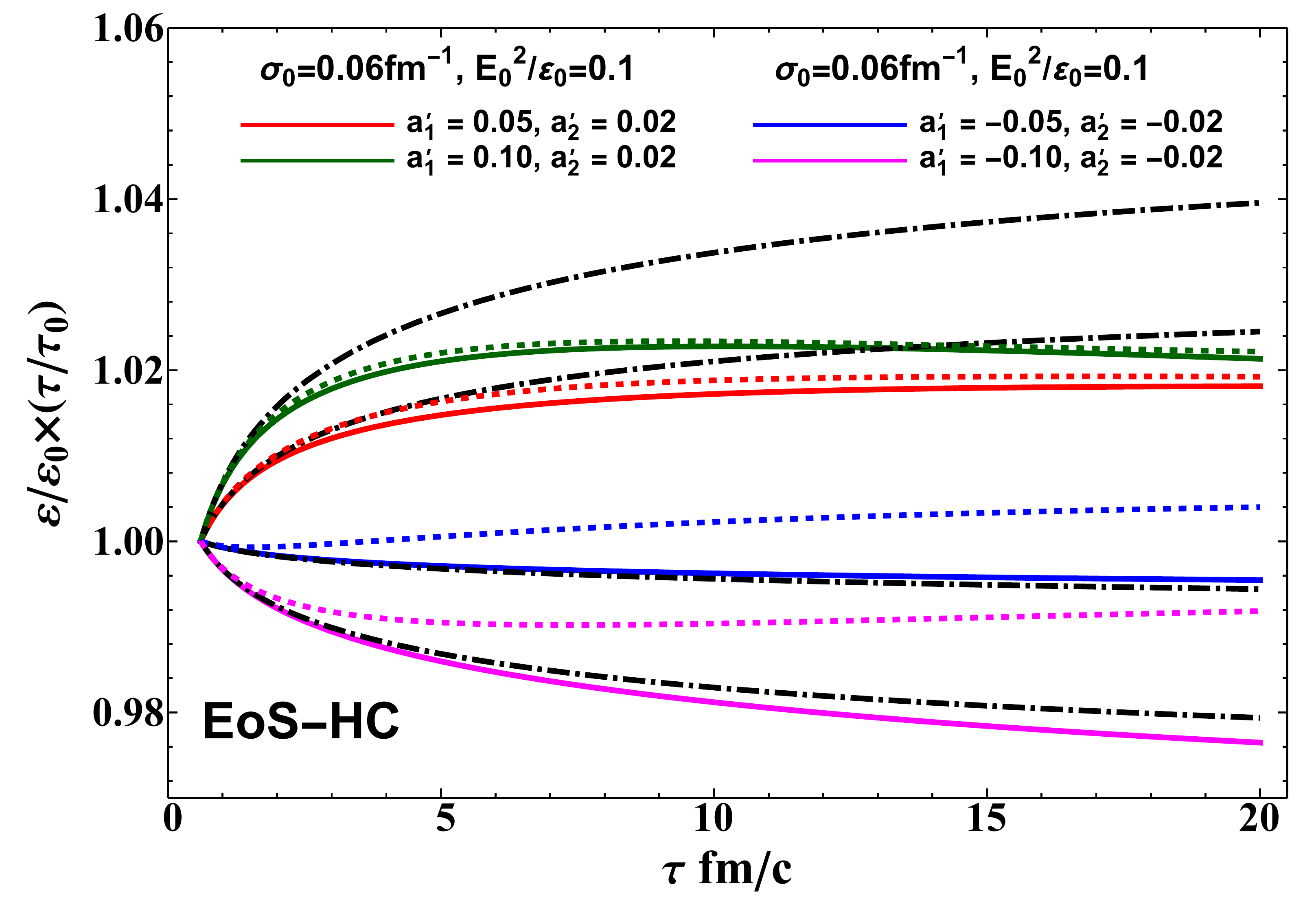}\includegraphics[scale=0.25]{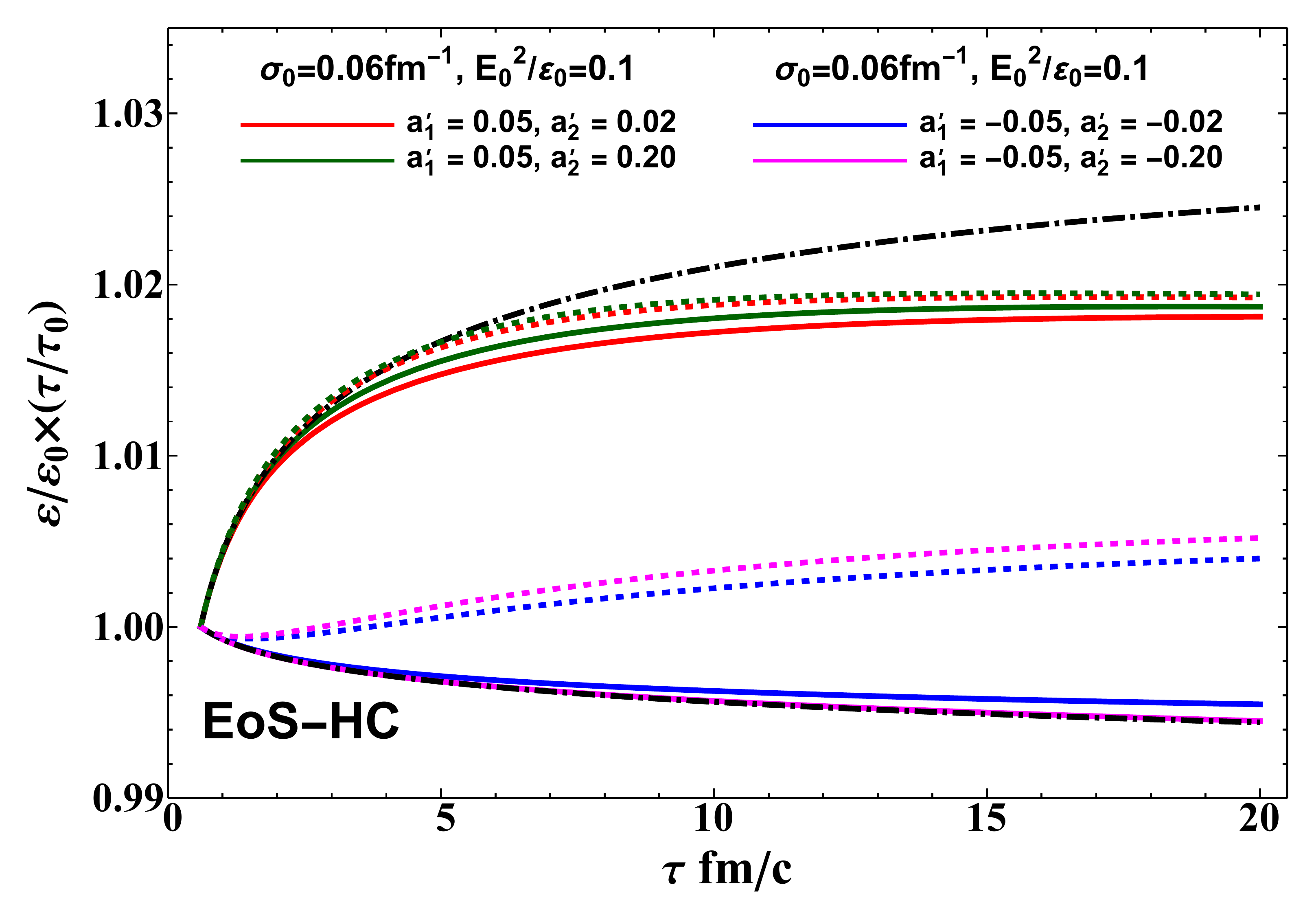}

\includegraphics[scale=0.25]{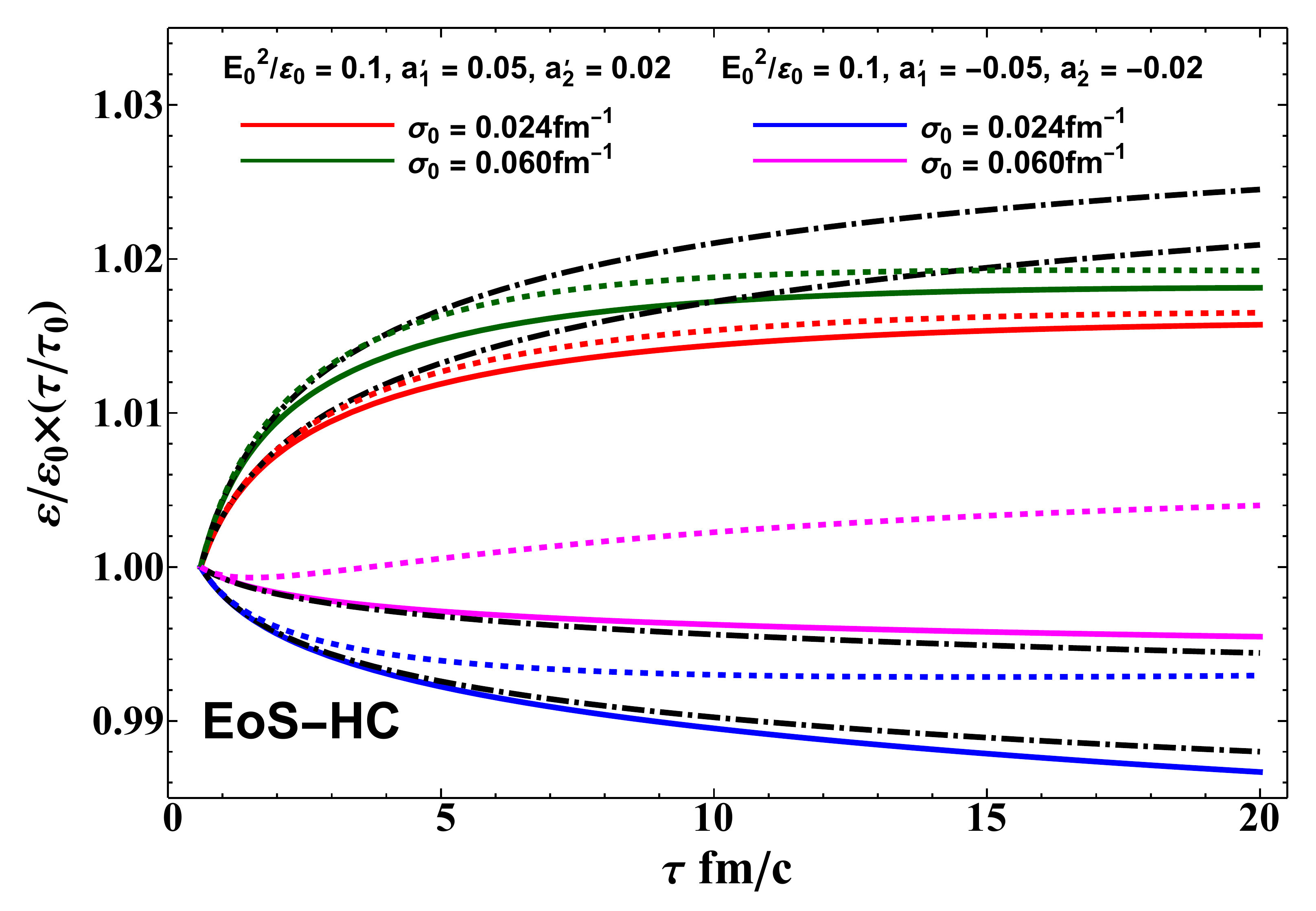}\includegraphics[scale=0.25]{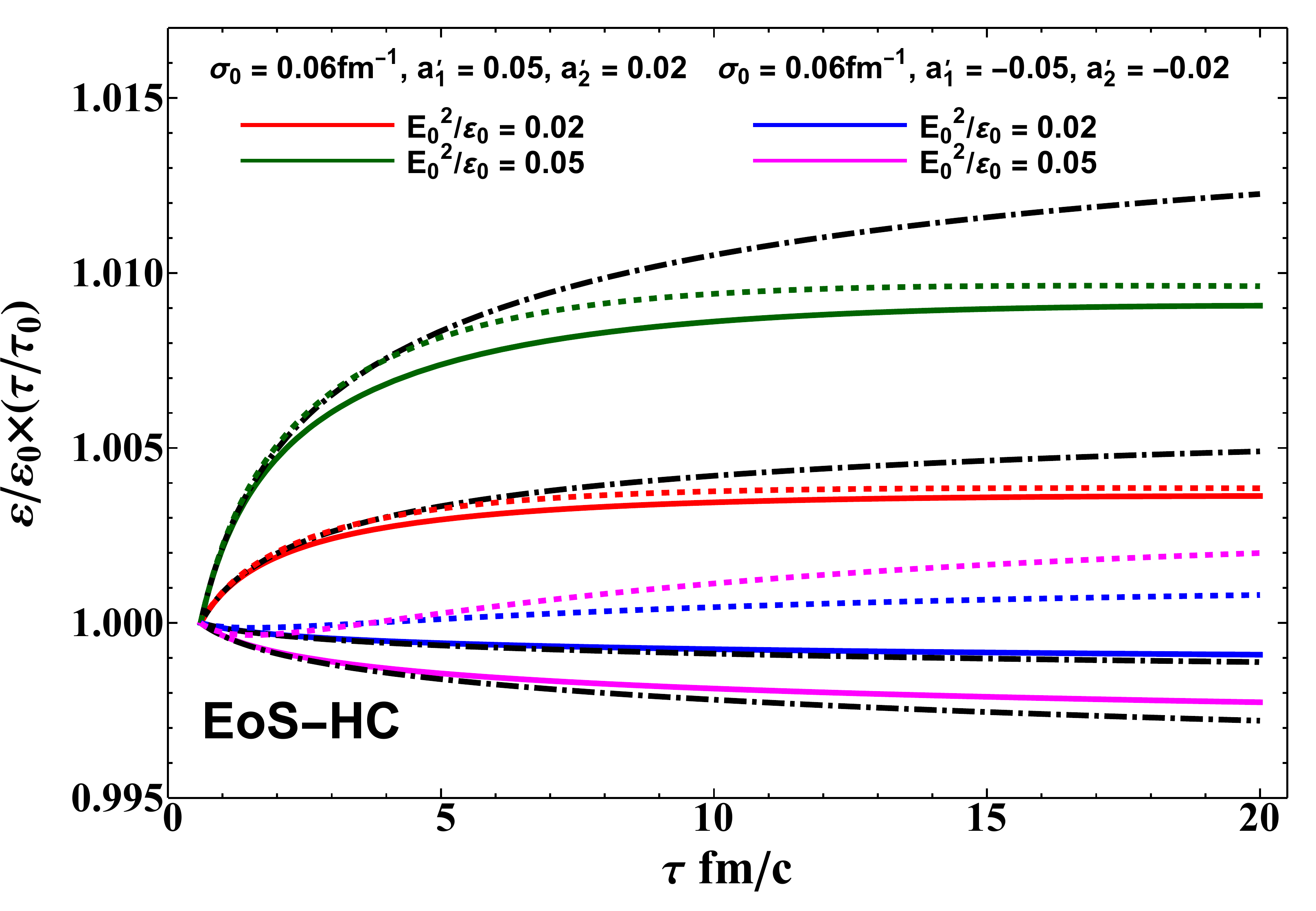}

\caption{The proper time scaled energy density $\varepsilon/\varepsilon_{0}\times\left(\tau/\tau_{0}\right)^{1+c_{s}^{2}}$
computed with EoS-HC (\ref{eq:EoS_HC}) as functions of proper time
$\tau$ with different parameters $a_{1},a_{2},\sigma_{0}$ and $E_{0}^{2}/\varepsilon_{0}$.
The colored solid, black dashed-dotted and colored dotted lines denote
the results for solving the results from Eqs. (\ref{eq:EN_01}, \ref{eq:EN_02})
numerically, analytic solutions (\ref{eq:EN_03}, \ref{eq:soln_xyz_02})
and numerical results with a constant $\sigma$ derived in Ref. \citep{Siddique:2019gqh,Wang:2020qpx},
respectively.\label{fig:en_p}}
\end{figure}

Following the same method in the previous section, we can get the
Eqs. (\ref{eq:EN_01}, \ref{eq:EN_02}, \ref{eq:EN_03}, \ref{eq:xyz_01}).
The conductivity (\ref{eq:Sigma_HC}) becomes, 
\begin{equation}
\sigma=\sigma_{0}\left(\frac{\tau_{0}}{\tau}\right)^{\frac{2}{3}-c_{s}^{2}}y^{2/3}\cdot z^{-c_{s}^{2}/(1+c_{s}^{2})}.\label{eq:Sigma_04}
\end{equation}
Using variables $x,y,z$, Eqs. (\ref{eq:EN_02}) becomes, 
\begin{eqnarray}
\frac{d}{d\tau}x & = & -\sigma_{0}\left(\frac{\tau_{0}}{\tau}\right)^{\frac{2}{3}-c_{s}^{2}}xy^{2/3}\cdot z^{-c_{s}^{2}/(1+c_{s}^{2})}-\frac{a_{1}^{\prime}}{\tau_{0}}\left(\frac{\tau_{0}}{\tau}\right)^{1/3}y^{1/3},\nonumber \\
\frac{d}{d\tau}y & = & a_{2}^{\prime}\frac{x}{\tau},\nonumber \\
\frac{d}{d\tau}z & = & \frac{E_{0}^{2}}{\varepsilon_{0}}\sigma_{0}\left(\frac{\tau_{0}}{\tau}\right)^{\frac{5}{3}-2c_{s}^{2}}x^{2}y^{2/3}z^{-c_{s}^{2}/(1+c_{s}^{2})}+\frac{a_{1}^{\prime}}{\tau_{0}}\left(\frac{E_{0}^{2}}{\varepsilon_{0}}\right)\left(\frac{\tau_{0}}{\tau}\right)^{\frac{4}{3}-c_{s}^{2}}xy^{1/3},\label{eq:Final_02}
\end{eqnarray}
where $x(\tau_{0})=y(\tau_{0})=z(\tau_{0})=1$, and dimensionless
constants $a_{1}^{\prime}$, $a_{2}^{\prime}$ read 
\begin{eqnarray}
a_{1}^{\prime} & = & eC\chi\left(\frac{n_{5,0}}{a}\right)^{1/3}\frac{B_{0}}{E_{0}}\tau_{0},\nonumber \\
a_{2}^{\prime} & = & e^{2}C\chi\frac{E_{0}B_{0}}{n_{5,0}}\tau_{0}.\label{eq:a123_02}
\end{eqnarray}

We follow the same power counting in Sec.\ref{sec:solution_HT}, i.e.
$\mathcal{O}\left(\frac{E_{0}^{2}}{\varepsilon_{0}}\right)\sim\mathcal{O}(a_{1,2}^{\prime})\sim\mathcal{O}(\hbar)$
and derive the analytic solutions up to the order of $\hbar$, 
\begin{eqnarray}
x(\tau) & = & e^{m_{1}^{\prime}(\tilde{\tau}^{\prime}-1)}+\frac{3}{2}a_{1}e^{m_{1}^{\prime}\tilde{\tau}^{\prime}}(n_{1}^{\prime}-1)[\mathcal{E}(n_{1}^{\prime},m_{1}^{\prime})-\tilde{\tau}^{\prime2/(1+3c_{s}^{2})}\mathcal{E}(n_{1}^{\prime},m_{1}^{\prime}\tilde{\tau}^{\prime})],\nonumber \\
y(\tau) & = & 1-\frac{3}{2}a_{2}e^{-m_{1}^{\prime}}(n_{1}^{\prime}-1)[\mathcal{E}(1,-m_{1}^{\prime})-\mathcal{E}(1,-m_{1}^{\prime}\tilde{\tau}^{\prime})],\nonumber \\
z(\tau) & = & 1-m_{1}^{\prime}\frac{E_{0}^{2}}{\varepsilon_{0}}e^{-2m_{1}^{\prime}}[\mathcal{E}(-2n_{1}^{\prime}+1,-2m_{1}^{\prime})-\tilde{\tau}^{\prime2n_{1}^{\prime}}\mathcal{E}(-2n_{1}^{\prime}+1,-2m_{1}^{\prime}\tilde{\tau}^{\prime})]\nonumber \\
 &  & -\frac{3}{2}a_{1}\left(\frac{E_{0}^{2}}{\varepsilon_{0}}\right)e^{-m_{1}^{\prime}}(n_{1}^{\prime}-1)[\mathcal{E}(-n_{1}^{\prime}+1,-m_{1}^{\prime})-\tilde{\tau}^{\prime n_{1}^{\prime}}\mathcal{E}(-n_{1}^{\prime}+1,-m_{1}^{\prime}\tilde{\tau}^{\prime})].\label{eq:soln_xyz_02}
\end{eqnarray}
where we introduce 
\begin{eqnarray}
n_{1}^{\prime} & = & \frac{-1+3c_{s}^{2}}{1+3c_{s}^{2}},\nonumber \\
m_{1}^{\prime} & = & -\frac{\sigma_{0}\tau_{0}}{\frac{1}{3}+c_{s}^{2}},\nonumber \\
\tilde{\tau}^{\prime} & = & \left(\frac{\tau_{0}}{\tau}\right)^{-\frac{1}{3}-c_{s}^{2}}.
\end{eqnarray}

For the numerical calculations, we choose the $\tau_{0}=0.6\textrm{ fm/c}$
and the speed of sound $c_{s}^{2}=1/3$. Meanwhile, we fix $E_{0}^{2}/\varepsilon_{0}=0.1$
for the discussion on $E/E_{0}\times\left(\tau/\tau_{0}\right)$ and
$n_{5}/n_{5,0}\times\left(\tau/\tau_{0}\right)$ again. 

In Figs. \ref{fig:EM_p}, we plot the proper time scaled electric
field $E/E_{0}\times\left(\tau/\tau_{0}\right)$, chiral density $n_{5}/n_{5,0}\times\left(\tau/\tau_{0}\right)$
as functions of the proper time $\tau$ with different parameters
$a_{1}^{\prime},a_{2}^{\prime}$ and $\sigma_{0}$. Interesting, in
the comparison with Fig. \ref{fig:EM_T}, the proper time scaled electric
field in small $|a_{2}^{\prime}|$ cases and chiral density, computed
with EoS-HC, seem to be close the those computed with EoS-HT when
we choose $a_{1}=a_{1}^{\prime}$ and $a_{2}=a_{2}^{\prime}$. It
means that the evolution of electric fields in the small $|a_{2}^{\prime}|$
cases and chiral density are not sensitive to the EoS.

Different with constant electric conductivity case in Ref. \citep{Siddique:2019gqh,Wang:2020qpx}
(colored dotted lines in Fig. \ref{fig:EM_p}), the temperature and
energy density dependent $\sigma(\tau)$ with EoS-HC will slow down
the decaying of electric fields when $a_{2}^{\prime}<0$. Such behavior
is also also insensitive to the $|a_{2}^{\prime}|$ when $a_{2}^{\prime}<0$. 

In Fig. \ref{fig:en_p}, we present the proper time scaled energy
density $\varepsilon/\varepsilon_{0}\times\left(\tau/\tau_{0}\right)^{1+c_{s}^{2}}$
as functions of the proper time $\tau$ with different parameters
$a_{1}^{\prime},a_{2}^{\prime},\sigma_{0}$ and $E_{0}^{2}/\varepsilon_{0}$.
We observe the similarity in small $|a_{2}^{\prime}|$ cases between
the figures in Fig. \ref{fig:en_p} and in Fig. \ref{fig:en_T}. When
$a_{2}^{\prime}<0$, the evolution of energy density is insensitive
to the $|a_{2}^{\prime}|$. The temperature and energy density dependent
electric conductivity $\sigma(\tau)$ will accelerate the decaying
of energy density when $a_{2}^{\prime}<0$. Combining with the results
in Fig. \ref{fig:EM_p}, it implies that the temperature and energy
density dependent electric conductivity $\sigma(\tau)$ helps the
electric fields gain the energy from the medium. 

In this section, we have solved the MHD with the temperature and energy
density dependent $\sigma(\tau$) in Eq. (\ref{eq:Sigma_HC}) and
EoS-HC. We present both the analytic solutions (\ref{eq:soln_xyz_02})
up to the $\mathcal{O}(\hbar)$ and the numerical results. 


\section{Splitting of global polarization for $\Lambda$ and $\overline{\Lambda}$
hyperons induced by magnetic fields \label{sec:magnetic-field-induced}}

In this section, we implement the results for relativistic MHD to
the global polarization of $\Lambda$ and $\overline{\Lambda}$ hyperons
in the relativistic heavy ion collisions. 

The polarization pesudo-vector is given by the modified Cooper-Frye
formula \citep{Becattini:2013fla,Fang:2016uds}, 
\begin{equation}
\mathcal{S}^{\mu}({\bf p})=\frac{\int d\Sigma\cdot p\mathcal{J}_{5}^{\mu}(p,X)}{2m_{\Lambda}\int d\Sigma\cdot\mathcal{N}(p,X)},\label{eq:S_01}
\end{equation}
where $m_{\Lambda}$ is the mass of $\Lambda$ hyperon, $\Sigma$
is the freeze out hypersurface, $\mathcal{J}_{5}^{\mu}(p,X)$ and
$\mathcal{N}^{\mu}(p,x)$ are axial and vector components of Wigner
functions in phase space. By inserting the general solutions in Wigner
functions \citep{Gao:2012ix,Chen:2012ca,Hidaka:2016yjf,Hidaka:2017auj,Hidaka:2018ekt,Hidaka:2022dmn,Fang:2022ttm},
the polarization vector $\mathcal{S}^{\mu}$ in Eq. (\ref{eq:S_01})
can be divide as several different parts \citep{Hidaka:2017auj,Yi:2021ryh,Yi:2021unq},
\begin{eqnarray}
\mathcal{S}^{\mu}(\mathbf{p}) & = & \mathcal{S}_{\textrm{thermal}}^{\mu}(\mathbf{p})+\mathcal{S}_{\textrm{shear}}^{\mu}(\mathbf{p})+\mathcal{S}_{\textrm{accT}}^{\mu}(\mathbf{p})+\mathcal{S}_{\textrm{chemical}}^{\mu}(\mathbf{p})+\mathcal{S}_{\textrm{EB}}^{\mu}(\mathbf{p}),
\end{eqnarray}
where 
\begin{eqnarray}
\mathcal{S}_{\textrm{thermal}}^{\mu}(\mathbf{p}) & = & \int d\Sigma^{\sigma}F_{\sigma}\epsilon^{\mu\nu\alpha\beta}p_{\nu}\partial_{\alpha}\frac{u_{\beta}}{T},\nonumber \\
\mathcal{S}_{\textrm{shear}}^{\mu}(\mathbf{p}) & = & \int d\Sigma^{\sigma}F_{\sigma}\frac{\epsilon^{\mu\nu\alpha\beta}p_{\nu}u_{\beta}}{(u\cdot p)T}\nonumber \\
 &  & \times p^{\rho}[\partial_{\rho}u_{\alpha}+\partial_{\alpha}u_{\rho}-u_{\rho}(u\cdot\partial)u_{\alpha}],\nonumber \\
\mathcal{S}_{\textrm{accT}}^{\mu}(\mathbf{p}) & = & -\int d\Sigma^{\sigma}F_{\sigma}\frac{\epsilon^{\mu\nu\alpha\beta}p_{\nu}u_{\alpha}}{T}\left[(u\cdot\partial)u_{\beta}-\frac{\partial_{\beta}T}{T}\right],\nonumber \\
\mathcal{S}_{\textrm{chemical}}^{\mu}(\mathbf{p}) & = & 2\int d\Sigma^{\sigma}F_{\sigma}\frac{1}{(u\cdot p)}\epsilon^{\mu\nu\alpha\beta}p_{\alpha}u_{\beta}\partial_{\nu}\frac{\mu_{B}}{T},\nonumber \\
\mathcal{S}_{\textrm{EB}}^{\mu}(\mathbf{p}) & = & 2\int d\Sigma^{\sigma}F_{\sigma}\left[\frac{\epsilon^{\mu\nu\alpha\beta}p_{\alpha}u_{\beta}E_{\nu}}{(u\cdot p)T}+\frac{B^{\mu}}{T}\right],\label{eq:S_all}
\end{eqnarray}
with 
\begin{eqnarray}
F^{\mu} & = & \frac{\hbar}{8m_{\Lambda}\Phi(\mathbf{p})}p^{\mu}f_{eq}(1-f_{eq}),\nonumber \\
\Phi(\mathbf{p}) & = & \int d\Sigma^{\mu}p_{\mu}f_{eq}.\label{eq:def_N}
\end{eqnarray}
where $\Phi(\mathbf{p})=\int d\Sigma^{\mu}p_{\mu}f_{eq}(x,p)$ is
the particle number density at freeze out hypersurface, $\mu_{B}$
is the baryon chemical potential and $f_{eq}$ is assumed as the standard
Fermi-Dirac distribution function. The $\mathcal{S}_{\textrm{thermal}}^{\mu}(\mathbf{p}),\mathcal{S}_{\textrm{shear}}^{\mu}(\mathbf{p}),\mathcal{S}_{\textrm{accT}}^{\mu}(\mathbf{p})$
for $\Lambda$ hyperons are the same as those for $\overline{\Lambda}$
hyperons. While, due to the difference in charge and baryon number,
$\mathcal{S}_{\textrm{chemical}}^{\mu}(\mathbf{p})$ and $\mathcal{S}_{\textrm{EB}}^{\mu}(\mathbf{p})$
flip their sign when we change $\Lambda$ to $\overline{\Lambda}$
hyperons. Therefore, the gradient of $\mu_{B}/T$ and EM fields can
lead to the splitting of global polarization for $\Lambda$ and $\overline{\Lambda}$
hyperons.

In order to compare with experiments, we need to transform the polarization
pseudo vector in the rest frame of $\Lambda$ and $\overline{\Lambda}$,
named $\vec{P}^{*}(\mathbf{p})$
\begin{eqnarray}
\vec{P}^{*}(\mathbf{p})=\vec{P}(\mathbf{p})-\frac{\vec{P}(\mathbf{p})\cdot\vec{p}}{p^{0}(p^{0}+m)}\vec{p},
\end{eqnarray}
where 
\begin{eqnarray}
P^{\mu}(\mathbf{p})\equiv\frac{1}{s}\mathcal{S}^{\mu}(\mathbf{p}),
\end{eqnarray}
with $s=1/2$ being the spin of the particle. Finally, the local polarization
is given by the averaging over momentum and rapidity as follows, 
\begin{eqnarray}
\langle\vec{P}(\phi_{p})\rangle=\frac{\int_{y_{\text{min}}}^{y_{\text{max}}}dy\int_{p_{T\text{min}}}^{p_{T\text{max}}}p_{T}dp_{T}[\Phi(\mathbf{p})\vec{P}^{*}(\mathbf{p})]}{\int_{y_{\text{min}}}^{y_{\text{max}}}dy\int_{p_{T\text{min}}}^{p_{T\text{max}}}p_{T}dp_{T}\Phi(\mathbf{p})},\label{eq:local_P_01}
\end{eqnarray}
where $\phi_{p}$ is the azimuthal angle. 

In this work, we concentrate on the splitting induced by the magnetic
fields. From Eq. (\ref{eq:S_all}), we introduce 
\begin{equation}
\Delta\mathcal{P}=\mathcal{P}_{\Lambda}-\mathcal{P}_{\overline{\Lambda}},
\end{equation}
where $\mathcal{P}_{\Lambda}$ and $\mathcal{P}_{\overline{\Lambda}}$
are the integration of local polarization (\ref{eq:local_P_01}) alone
the out-of-plane direction over the $\phi_{p}$. Taking $1-f_{eq}\simeq1$
in Eqs. (\ref{eq:S_all}), the $\Delta\mathcal{P}$ induced by the
magnetic fields is given by, $\Delta\mathcal{P}_{\textrm{EB}}\propto B^{y}/(m_{\Lambda}T)$.
Recalling the definition spin magnetic moment, following Ref. \citep{Muller:2018ibh},
$\Delta\mathcal{P}$ induced by magnetic fields is estimated as
\begin{equation}
\Delta\mathcal{P}_{\textrm{EB}}\approx2\frac{\mu_{\Lambda}}{T}\overline{B},\label{eq:P_EB}
\end{equation}
where $\mu_{\Lambda}=-0.613\mu_{N}$ is the spin magnetic moment for
$\Lambda$ hyperons. In the Ref. \citep{Muller:2018ibh}, $\overline{B}$
is averaged the magnetic field. Also see other related theoretical
studies on the splitting of polarization induced by EM fields \citep{Guo:2019joy,Buzzegoli:2022qrr,Xu:2022hql}.

Next, we will evaluate the $\Delta\mathcal{P}_{\textrm{EB}}$ in the
framework of relativistic magnetohydrodynamics. Recalling the expression
in Eq. (\ref{eq:local_P_01}), since the polarization pseudo vector
is computed near the freeze out surface, it is natural for us to choose
the $B^{y}(\tau_{f})$ with $\tau_{f}$ being the proper time for
the chemical freeze out instead of the averaged $\overline{B}$. For
a given collision energy, we choose the maximum value of space-averaged
initial magnetic fields in different impact parameters, which is computed
by Ref. \citep{Siddique:2021smf}. The initial proper time is chosen
as $\tau_{0}\simeq0.6$fm/c. The evolution of magnetic fields is given
by Eq. (\ref{eq:B_field}), which is also found in ideal MHD Ref.
\citep{Pu:2016ayh,Roy:2015kma,Pu:2016bxy,Pu:2016rdq}. For simplicity,
we choose the $\tau_{f}=10$fm/c for all collision energies. The temperature
for the chemical freeze out is followed by the experimental studies
in Ref. \citep{STAR:2021iop}. We summarize these parameters in different
collision energies in Tab. \ref{tab:T_B}.

\begin{table}
\caption{Collision energies, temperature given by Ref. \citep{STAR:2021iop}
, the initial space-averaged magnetic fields from Ref. \citep{Siddique:2021smf},
the magnetic fields estimated by Eq. (\ref{eq:B_field}) with $\tau_{f}=10$fm/c.
\label{tab:T_B}}

\centering{}%
\begin{tabular}{|c|c|c|c|c|c|c|}
\hline 
$\sqrt{s_{NN}}$ (GeV) & $7.7$  & $11.5$ & $27$ & $39$  & $62.4$  & $200$\tabularnewline
\hline 
\hline 
$T$ (MeV) & $144.3$ & $149.4$ & $155.0$ & $156.4$ & $160.3$ & $164.3$\tabularnewline
\hline 
space-averaged $eB(\tau_{0})/m_{\pi}^{2}$ & $9.1\times10^{-2}$ & $1.7\times10^{-1}$ & $4.5\times10^{-1}$ & $4.5\times10^{-1}$ & $3.2\times10^{-1}$ & $5.6\times10^{-2}$\tabularnewline
\hline 
$eB(\tau_{f})/m_{\pi}^{2}$ & $5.5\times10^{-3}$ & $1.0\times10^{-2}$ & $2.7\times10^{-2}$ & $2.7\times10^{-2}$ & $1.9\times10^{-2}$ & $3.3\times10^{-3}$\tabularnewline
\hline 
\end{tabular}
\end{table}

\begin{figure}
\begin{centering}
\includegraphics[scale=0.35]{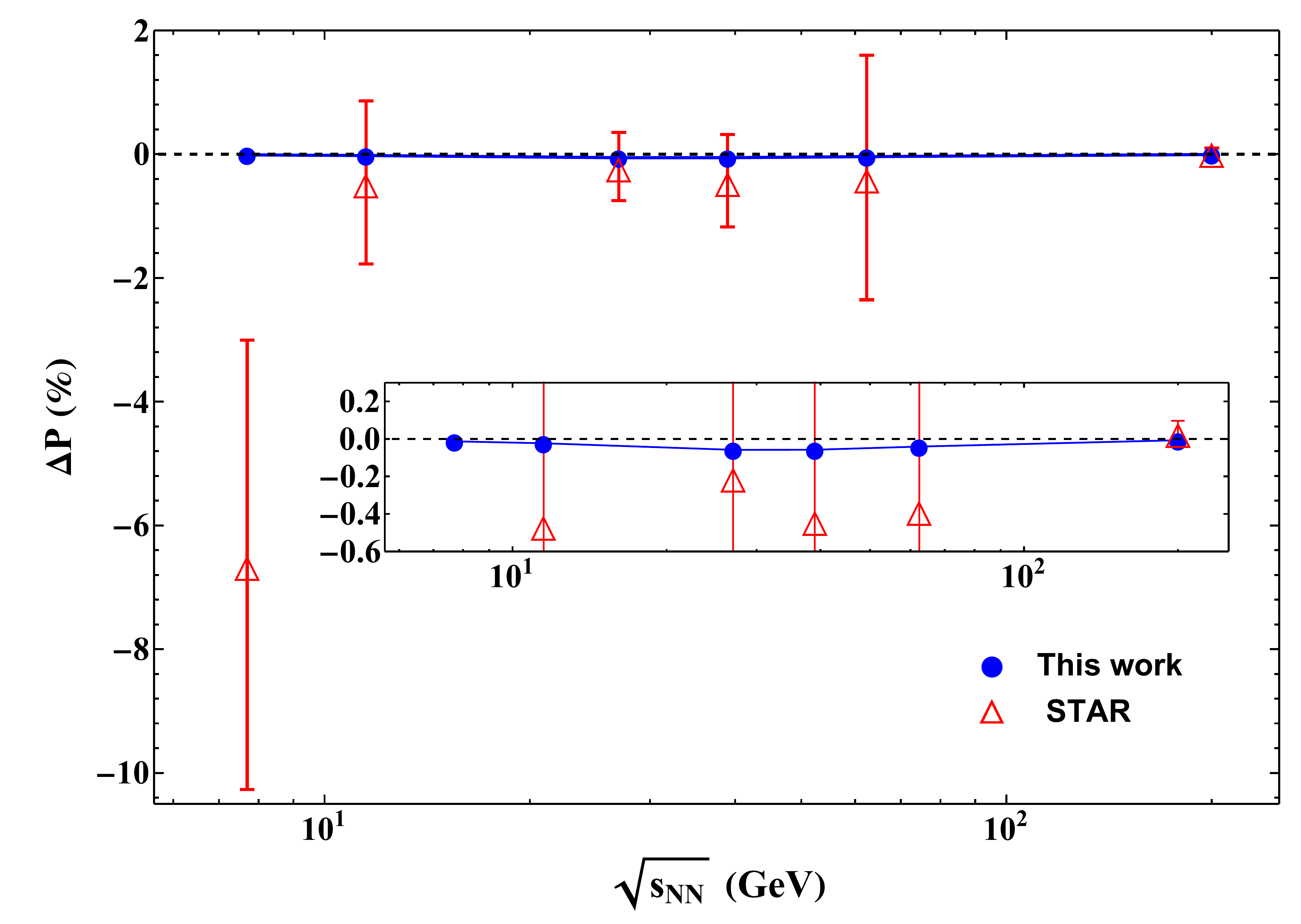}
\par\end{centering}
\caption{The difference of the global polarization between $\Lambda$ and $\overline{\Lambda}$
hyperons in different collision energies. The triangle points are
derived from STAR measurements \citep{STAR:2017ckg,STAR:2018gyt}.
The filled circle points denotes the estimation from Eqs. (\ref{eq:P_EB})
where the initial magnetic fields at $\tau_{0}=0.6$fm/c are given
by Ref. \citep{Siddique:2021smf} and the evolution of magnetic fields
are given by Eq. (\ref{eq:B_field}) with final time $\tau_{f}=10$fm/c.
\label{fig:deltaP}}
\end{figure}

In Fig. \ref{fig:deltaP}, we plot the the splitting of the global
polarization for $\Lambda$ and $\overline{\Lambda}$ hyperons, and
compare our estimation from Eq. (\ref{eq:P_EB}) with the data from
STAR measurements \citep{STAR:2017ckg,STAR:2018gyt} at $\sqrt{s_{NN}}=7.7,11.5,27,39,62.4,200\ \text{GeV}$
collisions. We find that the $\Delta\mathcal{P}_{\textrm{EB}}$ computed
from our framework agrees with the data expect the results at $\sqrt{s_{NN}}=7.7$
GeV. Roughly speaking, in the high energy collisions, the quark pairs
are generated from the vacuum and therefore, the net baryon number
and baryon chemical potential are approximately vanishing. The splitting
induced by EM fields may dominate in the global polarization of $\Lambda$
and $\overline{\Lambda}$ hyperons in the high energy region. On the
other hand, the collisions in low energy region, the net baryon density
is not negligible and will play a crucial role to the global and local
polarization, e.g. see Ref. \citep{Ryu:2021lnx} and the discussions
on the simulations for spin Hall effects in heavy ion collisions \citep{Fu:2022myl,Wu:2022mkr}.
Our results indicates that the magnetic fields may be not strong enough
to cause such huge splitting of global polarization in low energy
region. It implies that the contributions from $\nabla\mu_{B}/T$
may dominate the splitting of global polarization in low energy collisions
\citep{Ryu:2021lnx,Wu:2022mkr}.


\section{Summary \label{sec:4 Discussion-and-Summary} }

We have derived the solutions of the relativistic anomalous magnetohydronamic
with longitudinal Bjorken boost invariance and transverse electromagnetic
fields in the presence of temperature or energy density dependent
electric conductivity. 

After a short review on the anomalous MHD in Sec. \ref{sec:2 Anomalous-magnetohydrodynamics},
we simplify the energy-momentum and charge currents conversation equations
coupled to the Maxwell's equations. To close the system, we introduce
two kinds of EoS, i.e. EoS-HT (\ref{eq:EoS_HT}) and EoS-HC (\ref{eq:EoS_HC})
correspond to the high temperature and high chiral chemical potential
limits, respectively. The electric conductivity is also parameterized
as Eqs. (\ref{eq:Sigma_HT}) and (\ref{eq:Sigma_HC}) for the EoS-HT
and EoS-HC, respectively. We assume that the initial conditions for
the system is a Bjorken velocity (\ref{eq:Bjokren_velocity_01}) with
initial EM fields in Eq. (\ref{eq:EB_02}). After some calculations,
we confirm that the Bjorken boost invariance holds during the evolution.
The main differential equations reduce to Eqs. (\ref{eq:Max_02},
\ref{eq:Con_Eqs}). 

Next, in Sec. \ref{eq:Sigma_HT}, we derive the perturbative analytic
solutions (\ref{eq:EN_03}, \ref{eq:Soln_xyz_01}) up to the order
of $\hbar$ for the simplified differential equations (\ref{eq:Max_02},
\ref{eq:Con_Eqs}) with the EoS-HT (\ref{eq:EoS_HT}) and temperature
dependent conductivity (\ref{eq:Sigma_HT}). We present the numerical
solutions for electric fields, chiral density and energy density in
Figs.\ref{fig:EM_T}, \ref{fig:en_T}. We find that the temperature
dependent $\sigma(\tau)$ will quicken up the decaying of electric
fields. While, the decaying of chiral density seems not to be sensitive
to the $\sigma(\tau)$. 

Similarly, we derive the perturbative analytic solutions (\ref{eq:EN_03},
\ref{eq:soln_xyz_02}) up to the order of $\hbar$ with the EoS-HT
(\ref{eq:EoS_HC}) and temperature dependent conductivity (\ref{eq:Sigma_HC}).
The numerical results for the proper time scaled electric fields,
chiral density and energy density are shown in Figs. \ref{fig:EM_p},
\ref{fig:en_p}. We find that the decaying of electric fields and
energy density in the small $|a_{2}^{\prime}|$ limit and chiral density
seem not be affected by the EoS when we choose $a_{1}=a_{1}^{\prime}$
and $a_{2}=a_{2}^{\prime}$. On the other hand, when $a_{2}^{\prime}<0$,
temperature and energy density dependent electric conductivity $\sigma(\tau)$
will slow down or accelerate the decaying of electric fields or energy
density, respectively. 

At last, we implement the results for relativistic MHD to the global
polarization of $\Lambda$ and $\overline{\Lambda}$ hyperons in the
relativistic heavy ion collisions in Sec. \ref{sec:magnetic-field-induced}.
The splitting of global polarization for $\Lambda$ and $\overline{\Lambda}$
hyperons is estimated by Eq. (\ref{eq:P_EB}). In Fig. \ref{fig:deltaP},
we plot the the splitting of the global polarization for $\Lambda$
and $\overline{\Lambda}$ hyperons, and compare our estimation from
Eq. (\ref{eq:P_EB}) with the data from STAR measurements at $\sqrt{s_{NN}}=7.7,11.5,27,39,62.4,200\ \text{GeV}$
collisions. The $\Delta\mathcal{P}_{\textrm{EB}}$ computed from our
framework agrees with the data in both high and intermediate collisions
energies and fails at $\sqrt{s_{NN}}=7.7$ GeV. It implies that the
contributions from other sources, e.g. $\nabla\mu_{B}/T$, may dominate
the splitting of global polarization in low energy collisions.
\begin{acknowledgments}
The authors would like to thank Qun Wang for helpful discussion. This
work is supported in part by the National Key Research and Development
Program of China under Contract No. 2022YFA1605500 and National Natural
Science Foundation of China (NSFC) under Grants No. 12075235 and 12135011. 
\end{acknowledgments}

\bibliographystyle{h-physrev}
\bibliography{qkt-ref-20221117}

\end{document}